\documentclass{revtex4}

\usepackage{graphicx}
\newcommand{\pdf}{PDF\ }
\newcommand{\pdfs}{PDFs}

\newcommand{\ba}{\begin{eqnarray}}
\newcommand{\ea}{\end{eqnarray}}

\def\simge{\ \lower -2.5pt\hbox{$>$} \hskip-8pt \lower 2.5pt \hbox{$\sim$}\ }
\def\simle{\ \lower -2.5pt\hbox{$<$} \hskip-8pt \lower 2.5pt \hbox{$\sim$}\ }

\def\beqn{\begin{eqnarray}}
\def\eeqn{\end{eqnarray}}
\def\as{\alpha_s}

\begin{document}
% You should use BibTeX and revtex.bst for references
\bibliographystyle{revtex}

% number  and your conference paper identification number on the
% title page in preprint mode. Multiple \preprint commands are allowed.
\preprint{YITP-SB-02-01}

%\begin{flushright}
%{YITP-SB-02-01}
%\end{flushright}

\title{SUMMARY: WORKING GROUP ON QCD\\ AND STRONG INTERACTIONS}
% Optional argument for running titles on pages
%\title[]{}



\author{Edmond L.\ Berger and Stephen Magill}
\affiliation{Argonne National Laboratory, 9700 S.\ Cass Ave., Argonne, IL 60439-4815}

\author{Ina Sarcevic}
\affiliation{Department of Physics, University of Arizona, Tuscon, AZ 85721}

\author{Jamal Jalilian-Marian, William B.\ Kilgore, Anna Kulesza
and Werner Vogelsang}
%\thanks{}
\affiliation{Brookhaven National Laboratory, P.O.\ Box 5000, Upton, NY 11973-5000}

\author{Robert V.\ Harlander}
\affiliation{CERN, CH-1211 Geneva 23, Switzerland}

\author{Ed Kinney}
\affiliation{Department of Physics, University of Colorado, Boulder, CO 80309}

\author{Richard Ball}
\affiliation{University of Edinburgh, Edinburgh, Scotland, UK}

\author{Brenna Flaugher, Walter Giele, Paul Mackenzie and Zack Sullivan}
\affiliation{Fermilab, P.O.\ Box 500, Batavia, IL 60510-0500}

\author{Csaba Balazs and Laura Reina}
\affiliation{Florida State University, Tallahassee, FL 42306-4350}

\author{Wu-Ki Tung}
\affiliation{Department of Physics and Astronomy, Michigan State University, East Lansing, MI 48824-1116}

\author{Nikolaos Kidonakis, Pavel Nadolsky and Fredrick Olness}
\affiliation{Physics Department, Southern Methodist University, Dallas, TX 75275-0175}

\author{George Sterman}
\affiliation{C.N.\ Yang Institute for Theoretical Physics, SUNY Stony Brook, Stony Brook NY 11794-3840}

\author{Stephen D.\ Ellis}
\affiliation{Department of Physics, University of Washington, Seattle, WA 98195-1560}


\date{\today}

\begin{abstract}
In this summary of the considerations of the  QCD working group at Snowmass 2001,
the roles of quantum chromodynamics in the Standard Model and in the search for
new physics are reviewed, with empahsis on frontier areas in the field.  We
discuss the importance of, and  prospects for, precision QCD in perturbative
and lattice calculations.  We describe new ideas in the analysis of parton
distribution functions and jet structure, and review progress in small-$x$ and
in polarization experiments.  
\end{abstract}

%\maketitle must follow title, authors, abstract and \pacs
\maketitle

%\section{}
%\label{}
%\subsection{}
%\subsubsection{}

\section{Introduction: The Languages of Quantum Chromodynamics}

Like Janus, quantum chromodynamics faces in a pair of directions.
It looks ahead, toward the high-energy frontier, as a source of, and
background to, new physics at short distances, and it looks behind, toward the
strong-coupling phenomena of confinement, hadronic structure and
chiral symmetry breaking.

We may think of QCD as an, perhaps as {\it the}, exemplary
quantum field theory.
It exhibits nearly all of the extraordinary
properties of quantum field theories that have been studied
within the past five decades \cite{tasi95,hqcd}.
These properties include its characteristic asymptotic
freedom at  short distances, confinement and chiral symmetry
breaking at long, as well as instantons, monopoles and related nonperturbative
vacuum structure.  Most important, all of these may be
studied in QCD in energy ranges currently available, and all must be
grappled with in any high energy experiment in which hadrons play
a role in either the initial or final state.
Out of strong interaction physics developed
the ideas of duality and strings, and now QCD serves
as an important testing-ground and paradigm for these now-mature
concepts.   It serves as a model
for nonperturbative analysis beyond Standard Model physics.
Quantum chromodynamics is,
through its spontaneous chiral symmetry breaking, the
origin of most of the mass  of ``bright" matter in the universe.
Finally, it is a
reservoir of  benchmark problems, including confinement,
and the description of its nonperturbative structure.

Quantum chromodynamics is a vast field, 
with subdisciplines, each characterized by
its own degrees of freedom,
including perturbative QCD, heavy quark
effective theory, nonrelativistic QCD,
chiral perturbation theory, lattice
QCD, discrete light cone QCD, not to mention nuclear physics.
In a sense, the great, nearly inexhaustible
challenge of strong interaction physics is to
delineate the applicability of these effective
pictures, to connect them to each other, and to
the underlying dynamics.
At present, we have no universal approach adequate
to treat all of these aspects of the theory.
We have instead a family of languages, appropriate
to different length scales.

\subsection*{{Perturbative QCD}}

The use of perturbation theory in QCD is based in
asymptotic freedom, in which the coupling of the theory
vanishes as the logarithm of the momentum scale at
which it is probed, 
$\alpha_s(Q) \sim 2\pi/\beta_0\ln(Q/\Lambda_{\rm QCD})$.
To use this property, however, we must identify observables that
depend upon the theory at some specified scale, $Q\gg \Lambda_{\rm QCD}$.
Such quantities, which are termed
``infrared safe",  take the general form
\begin{eqnarray}
\hspace{-20mm}
Q^2\sigma_{\rm SD}(Q,m) = {\sum}_n a_n\left({Q^2/\mu^2}\right)\,
\alpha_s^n(\mu)
+ {\sum}_p {1\over Q^p} A_p\left({Q/ \mu},\alpha_s(\mu)\right)\, .
\label{P5_expand}
\end{eqnarray}
Examples of infrared safe observables are jet,
event shape and total
cross sections in $\rm e^+e^-$ annihilation and weak
vector boson decay, for which $Q$ is the
total center-of-mass energy scale
and the vector boson mass, respectively.
The parameter $\mu$ is the renormalization
scale, which may be chosen in these cases equal to $Q$.
We have exhibited as well power corrections in $Q$,
which are primarily nonperturbative in origin.
We will encounter the important phenomenological
role of power corrections below.  The coefficients
$a_n$ may be calculated exactly.  They provide,
however, at best an asymptotic series for the observable
in question.

Despite their
variety, cross sections that are infrared safe
are the exception rather than the rule at high energy,
and are limited to those with purely leptonic initial states.
The range of perturbative methods is essentially extended by
identifying cross sections with large momentum
transfer which, although not directly infrared safe, exhibit the
factorization
of perturbative short-distance from nonperturbative long-distance
dynamics.  Usually, such a factorization is manifested in the
form of a convolution, for example, an integral over parton
momentum fraction,$``x"$.  Factorizations, in turn, imply
evolutions, or  resummations of logarithmic corrections.

In a simplified example, suppose that we can write a
``physical" cross section describing momentum transfer $Q$,
as a product
of short-distance and long-distance factors, which
we may think of as coefficient functions and parton distribution
functions (PDFs),
respectively:
\begin{eqnarray}
\hspace{-10mm}
Q^2\sigma_{\rm phys}(Q,m) &=& C_{\rm
SD}\left({Q\over\mu},\alpha_s(\mu)\right)\,
   f_{\rm LD}\left({\mu\over m},\alpha_s(\mu)\right)
+ \sum_n (1/Q^n)\, C_n\, ,
\label{P5_fact}
\end{eqnarray}
where the ``factorization scale" $\mu$ separates
long- and short-distance dynamics.  The nonleading, $C_n$
terms, suppressed by powers of the momentum transfer,
generally possess related, but more complex factorization
properties.
Since the choice of $\mu$ is arbitrary,
the $\mu$-dependence is determined to be power-like,
\begin{eqnarray}
\mu d\sigma_{\rm phys}/d\mu =0
   &\Rightarrow& \quad C\sim Q^{\delta(\alpha_s(\mu))} \quad f \sim
\mu^{\delta(\alpha_s(\mu))}\, .
\label{P5_evol}
\end{eqnarray}
The power, an ``anomalous dimension", can depend only upon
$\alpha_s(\mu)$, because this is the only scale held in
common by the two functions.  Then, because
the coefficient function can depend upon $Q$ only through
the ratio $Q/\mu$, the anomalous dimension controls
the dependence of the physical cross section on the momentum transfer.
Essentially all the evolutions
and resummations of perturbative QCD may be understood in
this language.  Perturbative QCD is not all of QCD, but the two
become equivalent for infrared safe quantities as $Q\rightarrow \infty$.

The formal basis of factorization is in the quantum-mechanical
incoherence of the long-distance
dynamics that binds hadrons from the 
the short-distance dynamics that governs large
momentum transfers, or the creation of extremely short-lived
virtual states (sensitive to new physics).  The product (or convolution)
forms that characterize factorized cross sections  express
this incoherence.  The probabilities for incoherent
processes may be multiplied, as if they were classical, and
may be calculated using different methods.  As long
as $\mu$ is large enough, however, perturbation theory may be
employed to control logarithmic behavior in the factorization scale.

The techniques of infrared safety, factorization and evolution,
involve a variety of expansions, in terms of
the coupling $\alpha_s$, of course,
but also, as we have seen in momenta $1/Q$.
In the presence of additional scales,
expansions in heavy quark masses, or
nonrelativistic velocities in units of $c$, are
appropriate in the same basic pattern.

\subsection*{{Lattice QCD}}

Sometimes thought of as the opposite extreme from perturbative QCD,
lattice QCD works best for static properties of the theory,
dependent on its long-distance structure.   Rather than
cross sections at large momentum transfer, its stock-in-trade
is operator expectation values, from which
may be determined  bound-state masses and, of increasing
interest,  matrix elements of local operators
such as electroweak currents.

A typical lattice expansion, analogous to Eq.\ (\ref{P5_expand}),
for the expectation of operator $J$, relates the physical, or
continuum expectation, to the corresponding matrix element in
the lattice-regularized theory.  This relation is of the general form,
\begin{equation}
\hspace{-20mm}
\langle J^{\rm cont} \rangle
   = Z\left(\mu a,\alpha_s(\mu)\right) \langle J^{\rm lat}(\mu a)\rangle
+
\sum_n a^n\, \langle K_n\left(\mu, a,\alpha_s(\mu)\right) \rangle\, ,
\label{P5_lattice}
\end{equation}
with $a$ the lattice spacing.
The matrix elements here should be understood to include the
nonlocal products of a few extra fields, through which expectations in
hadronic states can be generated.  In contrast to perturbative
methods, the expectation on the right-hand side, $\langle J^{\rm lat}(\mu
a)\rangle$
may be computed numerically with high accuracy on the lattice.
Numerical and finite-size issues aside,
the challenge in this case is that the lattice theory is not the continuum
theory, but is related to it by the limit of vanishing
lattice spacing, $a\rightarrow 0$.

The relation between the continuum and lattice matrix elements
bears a strong resemblence to the factorized cross section
in Eq.\ (\ref{P5_fact}), with the $Z$-factor playing the role
of the short-distance coefficient function, now a
function of $a$ in place of $Q$, and the expectation
value playing the role of the long-distance matrix element.  As in
the previous case, and indeed even more so in the lattice relation,
contributions that are power-suppressed in $a \leftrightarrow 1/Q$ play a 
crucial role.
Unlike a perturbative calculation, however, the output of the
lattice calculation is the long-distance matrix element itself.
The perturbative expansion of the $Z$-factor remains of
importance to make the transition, and control
over the expansion in $a$ requires special care.   In
realistic cases, quark masses also matter, and
we may variously encounter expansions in inverse powers of
heavy quark masses, and in (chiral) logarithms of light quark masses.

\subsection*{{Field-theoretic consistency}}

Perturbative and lattice QCD are ``first-principle" methods,
exact up to corrections that are relatively easy to
characterize, but are generally quite difficult to estimate.
Not all valuable information is found in this fashion, however.
The self-consistency of quantum chromodynamics itself, and of
QCD with the remainder of the Standard Model, imply
strong constraints on experiment, and provide powerful
frameworks with which to study the theory.

 From the consistency of perturbative QCD, supplemented by
the operator product expansion, its nonperturbative
spectrum and transition amplitudes, come
QCD Sum Rules.   Variants of this approach involve
such concepts of very current interest as the local duality
between partonic and hadronic degrees of freedom.
 From the necessary harmony of QCD with the electroweak
content of the Standard Model, including the Higgs
phenomenon, may be derived chiral perturbation theory,
and its very successful description of the lowest-energy
strong interactions.

Our understanding of the spectroscopy of QCD is still far from
complete, and efforts to move the frontier forward rely upon all of these methods,
from classic quark model concepts to the most mathematical
sleights of hand of duality.  Light-cone methods offer another synthesis of
perturbative and lattice ideas \cite{sjbsnow}.

\subsection*{QCD at Snowmass 2001}

The QCD Working Group (P5) was convened by Brenna Flaugher, Ed Kinney,
Paul Mackenzie and George Sterman.  The results of the five
topical subgroups are summarized in this report.  With their conveners, they were:
A. Parton distributions, spin and resummation (Walter Giele and Fred Olness),
B. Fixed-order perturbation theory and top (Bill Kilgore), C. Jets and jet algorithms
(Steve Ellis and Brenna Flaugher), D. Diffractive and nuclear QCD (Ina Sarcevic)
and E. Lattice QCD, light and heavy hadrons (Paul Mackenzie).
George Sterman prepared the general
sections of this report.  

It is not possible below to summarize all the ideas that were
discussed nor all the progress that was made at
Snowmass, some of which is surely still latent in the minds
of participants, and QCD is far too large a field to
address fully even in an extended workshop.  At the same time,
QCD issues were so integral to the considerations of 
other Physics, Experimental and Accelerator working groups,
that demands regularly outpaced supply, even for the deep intellectual  
resources assembled on the mountainside.
For example, stimulating discussions were held on very high energy
neutrino cross sections \cite{Reno01}, on the CESR and Jlab programs \cite{Dzie01} in
spectroscopy and on the interplay of chiral perturbation
theory with lattice QCD, which we will not touch on in detail below.
Another exciting but neglected area, closely related to those discussed
at Snowmass and below, regards elastic scattering and rare B decays.
Despite these omissions,
we hope that this Summary may be successful
in communicating at least some of the sense of breadth and excitement
of the meeting and of the field.

\section{Precision and a New Era of QCD}

In a sense, we may think of quantum chromodyanmics as a vast
continent, separating two oceans that represent its asymptotic
limits: one to ``high-$Q$" partonic degrees of freedom, the
other to ``low-$Q$" hadronic degrees of freedom.
Lattice methods address the latter most directly; perturbative
methods the former.  
Pursuing this
metaphor, the aim of resummations, as illustrated by
(\ref{P5_evol}), is to extend these analyses into a region
where the different pictures, in terms of different
degrees of freedom overlap.  The increased precision
for which we strive in computing higher-order perturbative
and power corrections in Eqs.\ (\ref{P5_expand})
and (\ref{P5_lattice}) is the route inland to this continent.
\begin{figure}[h]
%\vbox{\vskip 1 true in}
\hbox{\hskip 1.5 cm
\includegraphics[width=25pc]{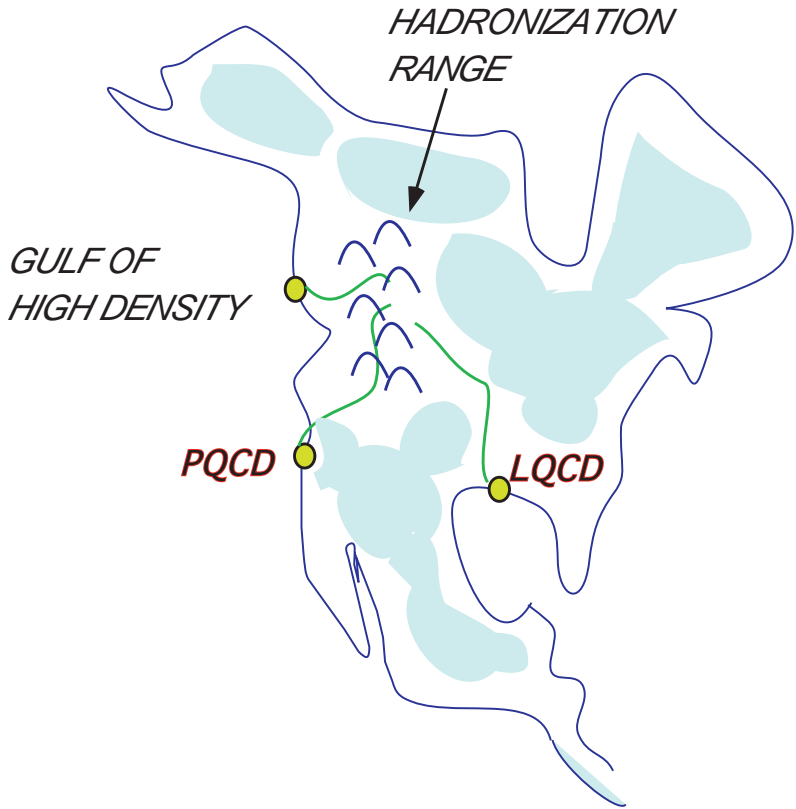}}
\caption{QCD as a continent.}
\label{continent}
\end{figure}

Increased precision leads to qualitative advances in
understanding when it requires the simultaneous treatment
of alternative degrees of freedom, the translation of
one language of QCD to another.
One example, now close at hand, is the role of hadronization
and confinement effects in jet cross sections, a problem
that is ``pure QCD".   The same principle of
alternative degrees of freedom  applies to the role of
QCD in the search of new physics.
The measurement of the strong coupling with an accuracy
necessary to extrapolate it beyond the
electroweak symmetry breaking scale will require precisely
this control over nonperturbative power corrections in
jet cross sections \cite{Abe:2001nq,burr01}.
Similarly, to detect physics at the TeV
scale that starts and mixes with quantum chromodynamics
will require an unprecedented ability to
interpret the strong-interaction structure of final states
\cite{atlas01,bauretal}.

Both QCD dyanmics and the search for new physics
require reliable extrapolations from long to short distances.
We may summarize this relationship with a variant of
Eq.\ (\ref{P5_fact}),
\begin{equation}
\sigma=C_{\rm SD,new+QCD}(Q,M_{\rm new},\mu)\,
\otimes\, f_{\rm LD,QCD}(\mu,m)\, .
\end{equation}
New physics at the scale $M$ ($\sim$ 1 TeV, say)  appears in the coeffcient
function, $C$, while the long distance function $f$
is controlled by QCD. In high-energy experiments that
involve direct production of new states, the momentum
transfer $Q\sim M$, and we will want to extrapolate
$f$ over a long lever arm in $Q/\mu$, from precision
experiments at current energies to the TeV range.
In experiments, such as those at B factories, where
we rely on rare events at fixed $Q$ ($\sim$ 5 GeV, say),
we need an equally precise understanding of the
long-distance function $f$, along with a careful
calculation of $C$ in a variety of new-physics scenarios.

The nineteen eighties saw the confirmation of the basic
ideas of quantum chromodynamics, which solidified its
place in the Standard Model.  
The past decade has seen a transformation in 
QCD, and the focus has moved from ``tests" of the
theory, toward its exploration.  
In a large sense, this step forward has been made
possible by a revolution in QCD data.
The multi-GeV
accelerator runs of the Tevatron, LEPI, SLC, HERA and LEPII
have provided data that exposed the partonic degrees
of freedom of QCD in an unprecedented fashion.  To
see the difference, one need only compare the 
illustrations from the {\it Review of Particle Properties}
from 1990 and 2000.

For this reason, as well as because of
advances in theory along many fronts, 
QCD is at the beginning of a new era.
This sense informed our discussions at Snowmass,
in a list of discussions
which could hardly have been imagined at the 
end of the 1980's, and  which may serve as
an outline for the remainder of this report. 

\begin{itemize}  

\item{} {From Next- to next-to-next \dots}  (Section \ref{sec::P5B}.) Two-loop
calculations for selected single-scale cross sections
have existed for some time.  The past year has seen 
the first calculations of true two-loop scattering
amplitudes.  These extraordinary new calculations
are still some way from implementation as factorized
cross sections, but the ice has been broken.  Two-loop
cross sections at large momentum transfer will make
possible percent accuracy in a wide range of
applications. 

On a related front, from the design of
detectors to the analysis of data, Monte Carlo event
generators play a central role in modern high energy physics.
These computer programs generally dress lowest-order 
partonic cross sections with parton evolution based upon
fragmentation, either independent or with QCD coherence
effects taken into account to some approximation.
The past few years has seen an increasing realization
that the full potential of forthcoming experiments cannot be reached 
without the use of the information that comes from
a more complete treatment of the hard scattering \cite{nloevent}.
The challenge is to do so without double-counting,
as a result of the definitions of the hard scattering
and fragmentation parts.  In a sense, this constitutes
the problem of how to generalize factorization
forms like Eq.\ (\ref{P5_fact}) to arbitrary numbers,
and quantum numbers,  of
outgoing particles.  There is no generally accepted
solution to this problem, but its great importance
to QCD phenomenology makes it a central part of the
long-term program in the field.

\item{} {Parton distribution uncertainties.} 
(Section \ref{sec:pdf}.) One of the
grand projects of QCD over the past thirty years has been
the determination of parton distribution functions.
Over the past few years, we have realized increasingly
the importance of estimating uncertainties for these
distributions, a complex enterprise involving estimates
of theoretical as well as experimental, and systematic
as well as statistical errors.  They will be an essential
ingredient in our analysis of new physics signals at
hadronic colliders.

\item{} Precision jet physics: measurements of $\alpha_s$.
(Section \ref{sec:jet}.)
Some of the most striking results of the highest energy data,
especially those from LEP, demonstrate the potential for a jet physics 
in which theory and experiment 
move from tens of percent toward the single percent level.
Such an improvement of precision will make possible determinations
of the strong coupling at a comparable level,
making in turn much more precise extrapolations
of the strong coupling, and hence of extensions of
the Standard Model.  It will stimulate an examination
of both perturbative and nonperturbative jet physics 
at a similar level, illustrating again the interdependence
of QCD physics and new physics searches.

\item{} Resummation phenomenology. 
(Section \ref{sec:resum}.) The reorganization of 
perturbative corrections, to fold
selected logarithmic contributions at all orders into
exact low order calculations, is beginning to provide
practical applications.  Extensive studies 
have been carried out for event shape
cross sections in $\rm e^+e^-$ annihilation, and 
in the transverse momentum distributions of electroweak
bosons produced through Drell-Yan processes in
hadronic collisions.  These may, in fact, be only the
beginning.

Closely related to resummation is the
analysis of power corrections, already begun 
for jet cross sections at LEPI and II, and at HERA.
The applications of resummation noted above require 
nonperturbative corrections, both theoretically
and phenomenologically.  

\item{} The dawn of unquenched lattice calculations.
(Section \ref{sec:lqcd}.)
Light quark loops are especially time-consuming in
lattice calculations, and the ``quenched approximation",
in which they are suppressed, has been heretofore a necessary evil 
for the majority of lattice results.  These results
can then reach an impressive precision, but they are
not QCD, even on a lattice.  Recent advances in lattice
theory, at least as much as improvements in computing 
power, have made it possible to extrapolate a series of
unquenched lattice calculations at the few percent level,
and a transition from single-hadron expectations to
multihadron amplitudes.
Such calculations can play a unique role in 
the analysis of  B decays, and the
determination of parameters of the CKM matrix.

\item{} Low-$x$, diffraction and gaps, high-density QCD.
(Section \ref{sec:smallxnuc}.)
The advent of high-energy nuclear scattering
is beginning to make it possible to study high-density
QCD in a partonic regime.  
The complementary program at HERA
maps out the long-distance partonic content of the proton,
at the edge of perturbative and nonperturbative 
dynamics.

\item{} {Polarized beams.} 
(Section \ref{sec:polarization}.) International programs 
are finding new windows into hadronic structure through
polarized beams.  Polarization capability can also provide 
sensitive tests of new physics, as shown at SLC.

\end{itemize}

\section{Fixed Order Calculations and Top Quark Physics}
\label{sec::P5B}

{\it This section was prepared by William B. Kilgore with input from
Robert V. Harlander,
Nikolaos Kidonakis,
Steve Magill,
Laura Reina and
Zack Sullivan.}

In the last several years, there have been a number of significant
developments in the field of computing higher order QCD corrections:
there have been technical developments in the computation of one-loop
and two-loop matrix elements, a number of general-purpose formalisms
for performing next-to-leading order Monte Carlo calculations have
been developed, and resummation techniques have been employed to
capture the most important components of higher order corrections to
processes for which it is currently unfeasible to perform a full
calculation beyond next-to-leading order.  Top quark physics continues
to be an active area of research with many important developments.

\subsection{NLO}
\label{sec::P5BNLO}
The process of performing next-to-leading order calculations is now
quite mature.  While the calculation of one-loop matrix elements is
still quite challenging, the loop integrals involved are well-known
and simplifying techniques (like the helicity method) for constructing
the amplitudes are available.  Once the matrix elements are known,
there are a variety of well-established methods for constructing
next-to-leading order Monte Carlo programs, including phase space
slicing~\cite{Giele:1992vf,Giele:1993dj}, the subtraction
method~\cite{Frixione:1996ms}\ and the dipole
method~\cite{Catani:1996jh,Catani:1997vz}.  A thorough review of these
methods and the various extensions to them has been made in
reference~\cite{Eynck:2001en}.

Studies of QCD observables have revealed that leading order
calculations are often little more than qualitative descriptions of
the processes in question.  While next-to-leading order corrections
generally change the prediction for the total rate of a process one
commonly finds that the corrections to the shapes of distributions of
observables are far more important than the change to the total
production rate.  When the signal is prominent relative to the
background, as in the case of massive vector boson production
(decaying to leptons) in association with jets, or even dominant, as
in the case of inclusive jet production, comparing measured
distributions with NLO predictions allows one to extract more precise
information about the process.

When backgrounds are large, however, the precise shapes of
distributions of observables may be needed just to extract the signals.
This observation points to a critical need for a much wider variety of
next-to-leading order calculations.  To date, the focus of such
calculations has understandably been on interesting signal processes.
However, it is expected that the distributions of background
processes also change when computed at higher order.  To some degree,
this problem is avoided by measuring the background in side-bands and
extrapolating into the signal region.  When the signal lies in a
steeply falling spectrum or at the tail of a background distribution,
however, it may be difficult to make an accurate extrapolation.  Thus,
it is increasingly clear that it is not enough to merely compute the
most important signal processes at NLO. The backgrounds to those
processes must also be computed at NLO if one is to extract the
interesting signal on the basis of shapes of distributions.

In some cases, like Higgs production and heavy flavor
production, leading order calculations severely underestimate the
cross section of the processes in question.  In such cases NLO
calculations are necessary but are probably no more reliable than one
would ordinarily expect leading order calculations to be and there is
a clear need to extend calculations to still higher order.

\subsection{NNLO}
\label{sec::P5BNNLO}
The developing state of the art in higher order corrections is
next-to-next-to-leading order (NNLO).  The first applications of NNLO
calculations will be to the signal processes with the most important
phenomenological applications.  Indeed, the first NNLO calculations
were to Drell-Yan production and the crossed process, deep inelastic
scattering.  In the last couple of years, the pieces have been
assembled to permit NNLO calculations of inclusive Higgs production,
dijet production in hadron-hadron collisions, $e^+e^-$ annihilation
into three jets and the crossed process, dijet production in deep
inelastic scattering.

To move beyond next-to-leading order QCD to next-to-next-to leading
order one needs two loop matrix elements.  Two rather different
techniques have been deployed recently.  For some time now, the
``integration by parts'' method has been used to compute multi-loop
vacuum polarization diagrams.  Baikov and Smirnov~\cite{Baikov:2000jg}
developed a technique to ``open-up'' the vacuum polarization diagrams,
permitting the application of the integration by parts method to
higher point amplitudes.  This technique was exploited by
Harlander~\cite{Harlander:2000mg} to compute the two-loop virtual
corrections to Higgs production by gluon fusion in the heavy top
limit.  With the virtual corrections available, it is now possible to
compute inclusive Higgs production at
NNLO~\cite{Harlander:2001is,Catani:2001ic,Harlander:2001eb}.

Another extremely important development has been the computation of
two-loop, four point scattering amplitudes.  The breakthrough came
with the solutions to the scalar double box master integrals for all
massless legs~\cite{Smirnov:1999gc,Tausk:1999vh} and the reduction of
tensor double boxes to scalar master
integrals~\cite{Smirnov:1999wz,Anastasiou:2000mf}.  With the integrals
in hand, it became possible to compute the two-loop amplitudes for
scattering two massless partons into two massless partons (the
amplitudes that contribute to single-jet-inclusive production) and
these have all been
computed~\cite{Anastasiou:2000kg,Anastasiou:2000ue,Glover:2001af}.

More recently, the master double box integrals with a single massive
external leg have been
solved~\cite{Gehrmann:1999as,Gehrmann:2000zt,Gehrmann:2001ck} and
applied to the calculation of the scattering amplitude which describes
such processes as $e^+e^-\to 3$ jets~\cite{Garland:2001tf}, dijet
production in deep inelastic scattering and high $p_T$ Drell-Yan
production.  The phenomenological application of these processes will
be profound.  An NNLO calculation of $e^+e^-\to 3$ jets will permit a
reanalysis of LEP data and a provide a much improved measurement of
$\alpha_s$.

In order to turn these two-loop amplitudes into NNLO calculations (for
hadronic collisions), one also needs NNLO parton distribution
functions (PDFs).  The most important source for extracting this
information is deep inelastic scattering (DIS).  While the DIS
coefficient functions have been known at NNLO for some
time~\cite{vanNeerven:1991nn,Zijlstra:1991qc,Zijlstra:1992qd,
Zijlstra:1992kj}, only partial results are currently available 
for the three-loop anomalous dimensions describing
parton evolution, in the
form of a finite number of fixed Mellin moments~\cite{Larin:1991zw,
Larin:1991tj,Larin:1997wd,Retey:2000nq,Gracey:1994nn,Bennett:1998ch,
Catani:1994sq,Fadin:1998py,Ciafaloni:1998gs}.  These partial results
have been used to construct approximations of the three-loop splitting
functions~\cite{vanNeerven:1999ca,vanNeerven:2000uj,vanNeerven:2000wp,
vanNeerven:2001pe} which, in turn, have been used to extract
approximate PDFs~\cite{Martin:2000gq,Alekhin:2001ih}.

To summarize the status of NNLO calculations: The Drell-Yan
process~\cite{Hamberg:1991np} (inclusive production of massive lepton
pairs) and the inclusive deep inelastic cross
section~\cite{Zijlstra:1992qd} have been known for approximately ten
years.  The inclusive production of scalar Higgs bosons has recently
been computed in the soft approximation and will soon be available for
the full kinematic region.  The matrix elements for the most important
$2\to2$ (or $1\to3$) scattering processes are now known.  The
splitting functions for computing three-loop parton evolution are not
yet fully known, but approximations based on a finite set of moments
have allowed approximate extractions of NNLO parton distribution
functions.  The algorithms for performing NNLO Monte Carlo
calculations have not yet been worked out and currently stand as the
largest barrier to the application of perturbative QCD at
next-to-next-to-leading order.

\subsection{Applications of Resummation to Fixed Order Calculations}
\label{sec::P5BResum}

For many interesting processes, it is not yet possible to perform
fixed order calculations beyond next-to-leading order.  Still one
would like to make use of the all-orders information available from
resummation techniques.  This is particularly true when one would like
to make measurements near a production threshold, where phase space
restrictions can lead to large logarithmic corrections.  These logs
occur at all orders in perturbation theory (beyond leading order) and
can lead to large corrections in the threshold region.  Thus, it is
important to capture these terms to obtain more accurate predictions
than NLO calculations can provide.
 
Threshold resummation~\cite{Berger:1995xz,Berger:1996ad,
Kidonakis:1996aq,Kidonakis:1997gm,Kidonakis:1998bk,Kidonakis:1998nf,
Laenen:1998qw} is performed in moment space and uses the factorization
properties of QCD to compute the large logarithmic corrections to all
order in $\alpha_s$.  To take this all-orders information back to
momentum space, one needs to invert the Mellin transform. This
inversion involves crossing the Landau pole of the QCD coupling and
requires a prescription. Unfortunately, different prescriptions lead
to different results for the sub-leading logarithms.  Without a direct
calculation beyond NLO, it is impossible to accurately determine the
sub-leading log terms.

A different technique is to expand the moment-space result in powers
of $\alpha_s$ and then transform the fixed order result to momentum
space, matching to the known NLO result.  This avoids prescription
dependence and unphysical terms and allows the accurate calculation of
leading and sub-leading logarithmic
terms~\cite{Kidonakis:2001nj,Kidonakis:2000ui}.  The current state of
the art in such calculations obtains results at NNLO-NNLL
(next-to-next-top-leading order and next-to-next-top-leading
logarithm)~\cite{Kidonakis:2001mc}.  Such analyses have recently been
applied to the inclusive jet production at NNLO-NLL and top quark
production at NNLO-NNLL~\cite{Kidonakis:2001mc}.

\subsection{Top Quark Physics}
\label{sec::P5BTop}
Top quark physics promises to be an active field of study in the
coming years, offering a variety of important measurements at Run II
of the Tevatron, at the LHC and at an $e^+e^-$ linear collider.  There
have been numerous surveys of top quark physics at various
colliders~\cite{Amidei:1996dt,Beneke:2000hk,Abe:2001nq} so here we
limit ourselves to developments presented at Snowmass 2001.

One of the most important measurements at the Tevatron and LHC will be
the top production cross section.  At the Tevatron, the cross section
is strongly affected by threshold corrections.  Threshold resummation
captures the most important logarithmic corrections, but is plagued by
prescription ambiguities.  By expanding the (moment-space) resummed
result in $\alpha_s$ before inversion to momentum space, ambiguities
can be avoided and one can obtain reliable predictions for the most
important logarithmic terms (up to next-to-next-to-leading log) at
NNLO~\cite{Kidonakis:1998bk,Kidonakis:1998nf,Kidonakis:2001mc}.

The associated production of a Standard Model Higgs boson, $h_{SM}$,
with a pair of top-antitop quarks can play a very important role at
the Tevatron and at the LHC, both for discovery and for precision
measurements of the top Yukawa coupling~\cite{Reina:2001et}. Recently
the total cross sections for $p\bar p\rightarrow t\bar th_{SM}$
\cite{Reina:2001sf,Reina:2001bc,Beenakker:2001rj} and $pp\rightarrow
t\bar th_{SM}$ \cite{Beenakker:2001rj} have been calculated at ${\cal
O}(\alpha_s^3)$, i.e. at next-to-leading order (NLO) in QCD.  The
calculation of both virtual and real ${\cal O}(\alpha_s)$ corrections
required the generalization of existing methods to the case in which
several external particles are massive.

The NLO QCD corrections slightly decrease the total cross section for
$p\bar p\rightarrow t\bar t h_{SM}$ at $\sqrt{s}\!=\!2$~TeV, when the
renormalization and factorization scales are varied between $m_t$ and $2
m_t$, while it increases it mildly for scales roughly above $2m_t$. On
the other hand, the impact of NLO QCD corrections on the total cross
section for $pp\rightarrow t\bar t h_{SM}$ at $\sqrt{s}\!=\!14$~TeV is
slightly positive over a broad range of scales.  More importantly, the
NLO results show a drastically reduced renormalization and
factorization scale dependence as compared to the Born result and
leads to increased confidence in predictions based on these results.

\section{Parton Distribution Functions (\pdfs)}
\label{sec:pdf}

{\it This section was
prepared with input from Csaba Balazs, 
Richard Ball, 
Edmond Berger,
Walter Giele,
Nikolaos Kidonakis, 
Pavel Nadolsky, 
Carlo Oleari,
Fred Olness,
Wu-Ki Tung and
Zack Sullivan.}

Reliable knowledge of parton distribution functions (\pdfs) is
required for detailed QCD studies and crucial
for many searches for new physics signals in the next generation of
experiments.  There has been significant recent progress in quantifying 
the uncertainties of the \pdfs.  The task is formidable and 
requires continuing study.  We briefly discuss these
issues, and examine prospects for improvement on both the
theoretical and experimental fronts~\cite{note}.

%%%%%%%%%%%%%%%%%%%%%%%%%%%%%%%%%%%%%%%%%%%%%%%%%%%%%%%%%%%%%%%%%%%%%%%%%%
\subsection{PDF Uncertainties: the Challenge} 
\label{sec:uncertainties}

Quantitative estimates of \pdf uncertainties are of great importance for 
current and future hadron collider experiments.  These uncertainties 
affect determinations of important quantities such as the $W$-boson mass, 
and they have an impact on the new physics potential of current and proposed 
hadron colliders.  Compositness searches in the Run Ia data on the one-jet 
inclusive transverse energy distribution~\cite{Abe:1996wy} are a case in 
point.  
Prior to this measurement, a series of global fits to PDFs, 
combining data from a variety of experiments
with next-to-leading order (NLO) perturbative QCD,  
had worked well, and were a showcase for the success of the 
QCD framework. 
Confronted with the relatively precise CDF one-jet inclusive measurement,
they produced
an apparent signal of quark compositness. However, these \pdfs\ could 
be adjusted to accommodate the high transverse
energy excess~\cite{Huston:1996tw}, 
while remaining consistent with previous data.  To many, this
demonstrated dramatically the need to
quantify uncertainties in \pdf determination at this level of
measurement accuracy. 
 
Quantifying \pdf uncertainties is difficult for many reasons:
1) From the practical side there are problems in combining
many different experiments in a consistent single global fit.
2) From the more theoretical side, one has to address many issues
to ensure correct error estimates, such as the
choice of \pdfs, validity of Gaussian approximations, likelihood
estimators, and other, equally technical considerations. 

In the last few years, much progress has been made to address these issues 
and obtain reliable \pdf sets that include uncertainties.
%Assuming the Gaussian approximation one can make a second order Taylor 
%expansion in \pdf parameter space around the minimum likelihood solution.
%Such an approach gives access to many analytic tools as the \pdf
%probability distribution is Gaussian. This method must ensure one has no
%directions in parameter space which have zero second derivative at the
%minimum likelihood solution. That is, the parameterization of the PDF
%functionals has to be finely tuned to on the one hand be able 
%to describe all experiments one fits to and on the other hand 
%restricted such that no zero modes exist. Apart from the Gaussian
%approximation this can lead to underestimating \pdf uncertainties.
%The first paper using this method ~\cite{Alekhin:1999za} used the H1,
%ZEUS, BCDMS and NMC data. Subsequently,
%refs.~\cite{Botje:2000dj,Barone:1999di} included additional experiments
%into the fits. In refs.~\cite{Stump:2001gu,Pumplin:2001ct} 
%a procedure to maintain the
%``global fitting'' philosophy by include as many experiments as
%possible was proposed.  To recover a probabilistic interpretation one has to
%multiply the experimental uncertainties with a common factor such that
%all experiments are consistent within a single \pdf ``global fit''.
For example, if a Gaussian probability distribution is assumed for the \pdfs,
systematic and statistical errors can be computed following general 
statistical
procedures based on the Bayesian treatment of uncertainties.  In this
approach, a covariance matrix of errors can be built by computing a second
order Taylor expansion in \pdf parameter-space around the maximum likelihood
solution.
This method is guaranteed to work 
(in the sense of a local minimum) if there are no directions in the parameter
space where the second derivative, computed with respect to the maximum 
likelihood 
solution, is zero.
In order to avoid this possibility, one has to finely tune the 
parameterization of 
the \pdfs, both to fit the data and to avoid zero modes.  
This method was first used in Ref.~\cite{Alekhin:1999za} to fit H1, ZEUS,
BCDMS and NMC data.  Other experiments were included in the fit in
Refs.~\cite{Botje:2000dj,Barone:1999di}, and in
Refs.~\cite{Stump:2001gu,Pumplin:2001ct} a procedure was proposed to maintain 
the ``global fitting'' philosophy, including as many experiments as possible.

%To recover a probabilistic interpretation one has to multiply the
%experimental uncertainties with a common factor such that all experiments are
%consistent within a single \pdf ``global fit''.

An alternative method developed in Ref.~\cite{Stump:2001gu} uses 
Lagrange multipliers to estimate \pdf uncertainties for a 
specific observable. Changing the 
observable necessitates re-fitting the \pdfs. However, using this method
one does not need to assume  a Gaussian distribution for the \pdfs.

A more general method that does not require a Gaussian approximation
is the Monte Carlo approach in which the space of \pdf functionals is sampled 
and probability weights are assigned to each \pdf
set~\cite{Giele:2001mr,Giele:2001ms}.
While the required computing resources are larger, one can in
principle incorporate any
error analysis. Moreover, the method can accommodate complicated topological
regions of \pdf parameter space that have a constant probability measure.
This approach opens the way to describe the \pdf functionals with a complete 
set 
of functions, with as many parameters as are needed numerically, and it 
removes 
the issue of parameterization dependence.  This method maximizes the
\pdf uncertainty and enhances the ability to consistently accommodate
a large number of functional forms.

\begin{figure}[h]
%\vbox{\vskip 1 true in}
\hbox{\hskip 3.5 cm
\includegraphics[width=25pc]{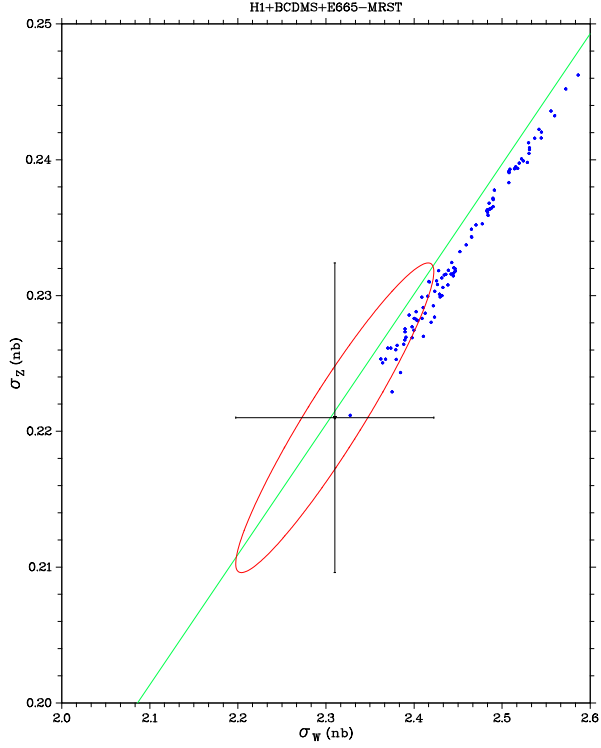}}
\caption{Scatter plot for \pdfs\ in the space of vector boson cross sections.
The D0 measurement is represented by the one-standard deviation error ellipse. 
The theory prediction is represented by 
a random sampling PDF set (using H1, ZEUS and E665).
Changing the luminosity moves the error 
ellipse over the diagonal line, changing the luminosity probability.}
\label{scatter}
\end{figure}

This strategy can be used to directly determine the luminosity at hadron
colliders from W-boson and Z-boson counts, as in Fig.\ \ref{scatter}.  In
Ref.~\cite{Giele:2001ms}, the luminosity  was derived from the
D0 Run I vector-boson events, and the result was compared with the
traditional method. Additionally, the corresponding implications for the top
quark cross section uncertainties were investigated using this extracted 
luminosity.  This study underscores the importance of this source of 
uncertainties, 
the understanding of which is crucial for a reliable luminosity determination 
and accurate
theoretical predictions.

%%%%%%%%%%%%%%%%%%%%%%%%%%%%%%%%%%%%%%%%%%%%%%%%%%%%%%%%%%%%%%%%%%%%%%%%%%
\def\fredhead#1{\null \vskip 0pt \noindent {\sc #1} \par}
\def\fredhead#1{\null \vskip 0pt \noindent {\sc #1}}

%%%%%%%%%%%%%%%%%%%%%%%%%%%%%%%%%%%%%%%%%%%%%%%%%%%%%%%%%%%%%%%%%%%%%%%%%%
\subsection{Projects at Snowmass} 
\label{sec:projects}

Having posed the problem of extracting \pdfs\  and their uncertainties, 
we briefly mention some the ongoing work that was performed during the 
Snowmass
workshop. Details can be found in the corresponding individual contributions
contained in these proceedings.

%%%%%%%%%%%%%%%%%%%%%%%%%%%%%%%%%%%%%%%%%%%%%%%%%%%%%%%%%%%%%%%%%%%%%%
\fredhead{PDF Standard User Interface}

In a continuation of the work initiated at 
the Les Houches Workshop~\cite{LesHouches},
a  standard user interface was presented and discussed. 
The purpose of this user interface is to provide 
a simple function call which gives access to all
\pdfs\  with or without uncertainties. The standard proposed  
has a number of improvements over previous standards such as the CERN PDFLIB.
Some features of the Les Houches standard are:
 1) The ability to easily access complete sets of \pdfs,  to facilitate 
calculations of the PDF uncertainties in various observables.
 2) The definition of \pdfs\ through external files. This makes the actual
interface code independent of the specific \pdf set used, and makes the
use of different sets within one analysis trivial.

%%%%%%%%%%%%%%%%%%%%%%%%%%%%%%%%%%%%%%%%%%%%%%%%%%%%%%%%%%%%%%%%%%%%%%
\fredhead{PDF uncertainties in $W$-Higgs production}

In Ref.~\cite{pavel}, the authors estimate the uncertainty in associated 
$W$-Higgs boson production at Run II
of the Tevatron due to imprecise knowledge of \pdfs.  
They use a method proposed by the CTEQ 
Collaboration~\cite{Pumplin:2001ct}
to estimate the uncertainties using a set of orthogonal PDF parameters,
obtained by the diagonalization of the matrix of second derivatives of
$\chi^2$ (Hessian matrix) near the minimum of $\chi^2$.
The PDF uncertainty for the signal and background
rates was found to be of the order of 3\%.
The uncertainties on important statistical quantities (significance of
Higgs boson discovery, accuracy of the measurement of the Higgs boson
cross section) are significantly smaller ($\sim$ 1.5\%) due
to the strong correlation of the signal and background. 
To summarize, the PDF error for the  $W$-Higgs production at Tevatron
is well under control.

%%%%%%%%%%%%%%%%%%%%%%%%%%%%%%%%%%%%%%%%%%%%%%%%%%%%%%%%%%%%%%%%%%%%%%
\fredhead{Heavy-quark parton distribution functions and their
uncertainties}

The uncertainties of the heavy-quark
($c$, $b$) \pdfs\  in the zero-mass variable flavor number scheme were 
investigated in Ref.~\cite{zack}.  Because
the charm- and bottom-quark \pdfs\  are constructed predominantly from
the gluon PDF, their uncertainties are commonly assumed to be the same as
the gluon uncertainty.  While this approximation is a reasonable first
guess, it was found to work better for bottom quarks than for charm quarks.  
The
heavy-quark uncertainties have a weak logarithmic dependence on $Q^2$ 
and approach the uncertainty of the gluon only for small $x$.  As an
application, the PDF uncertainty for $t$-channel single-top
production is calculated, and a cross section of $2.12 + 0.32 -
0.29$~pb at Run II of the Tevatron is predicted.

%%%%%%%%%%%%%%%%%%%%%%%%%%%%%%%%%%%%%%%%%%%%%%%%%%%%%%%%%%%%%%%%%%%%%%
\fredhead{Differential Distributions for NLO  Neutrino-Production of Charm}

In Ref.~\cite{kmo},
the charged current DIS charm production cross section is computed at NLO 
in QCD.   
While the inclusive calculation for this process has been published
previously, the fully differential distribution is necessary to properly
model the experimental detector acceptance. This full distribution, in turn, 
is crucial for correctly extracting the strange-quark PDF from the data. 
This work is a collaborative effort between theorists and experimentalists 
from the NuTeV collaboration, and this method will be implemented in the
analysis of the NuTeV di-muon data set. 

%%%%%%%%%%%%%%%%%%%%%%%%%%%%%%%%%%%%%%%%%%%
\subsection{Summary}

In recent years, new information has become available concerning parton
distributions and their uncertainties.  The issues have become more important
with the realization that these uncertainties could be hampering searches for
physics beyond the Standard Model.  To achieve the goal of quantitative PDF
uncertainties will require a comprehensive and collaborative investigation
involving both theorists and experimentalists.  
If we are to make the best use of data and discoveries from
hadron experiments, the area of PDF uncertainties must receive high priority.

%%%%%%%%%%%%%%%%%%%%%%%%%%%%%%%%%%%%%%%%%%%%%%%%%%%%%%%%%%%%%%%%%%%%%%%%%%

\section{Jet Physics}
\label{sec:jet}

{\it This section was written by Steve Ellis.}

As already noted, one of the most instructive arenas for the exploration of
QCD has been the study of jets in hadronic finals states. \ The underlying
picture is that a parton, which is isolated in momentum space due to some
short-distance hard scattering process, evolves via higher order corrections,
showering and hadronization to its long-distance asymptotic state consisting
of a ``spray'' or jet of hadrons, which is approximately aligned with the
direction of the initial parton. \ By assigning these associated hadrons to a
single jet, \textit{i.e}., by summing over their kinematic properties, one
constructs a measure of the hadronic final state that is infrared finite and
that tracks the kinematic properties of the underlying short-distance partons.
\ Thus 
jet cross sections
in $e^{+}e^{-}$ collisions can be directly
addressed in QCD\ perturbation theory. \ In the case of lepton-hadron or
hadron-hadron collisions, jet ideas can be used to control potential infrared
singularities arising due to the final state 
interactions, while
factorization techniques 
are
employed to control the singularities in the
initial state. \ Thus, much like the various shape parameters employed so
successfully to characterize $e^{+}e^{-}$ final states, \textit{e.g}., Thrust,
Energy-Energy Correlations, \textit{etc}., jet cross sections were first used
as tests of our understanding of QCD. \ As the physics focus has shifted, as
described elsewhere in this Summary, to the question of precision strong
interactions and new physics searches, the role of jet physics has also
shifted. \ The study of jets is not only central to precision
QCD\ 
measurements, but also 
essential as a tool for mapping the observed
long-distance hadronic final states onto the underlying short-distance
partonic states. \ While we get to measure only the former, it is the latter
that we can must recognize 
as the W and Z bosons and heavy 
quarks, which
we expect to arise from Higgs and Supersymmetric physics. \ A central element
of jet physics is the ``jet algorithm'', which is the set of rules that
specify how to identify and combine the hadrons (on the experimental side) or
the partons (on the theoretical side) into jets. \ As our control of higher
order corrections, resummations and nonperturbative 
contributions and
parton distributions improves, as outlined elsewhere in this Summary, we are
faced also with the task of improving the way in which we define jet
physics \cite{workshops}. \ 

Much of the discussion of jet physics at this Workshop focused on 
improving the definitions
of jets. \ Traditionally, two types of jet
algorithms have been employed at colliders. \ Both are based on the assumption
that hadrons associated with a jet will in some sense be nearby each other.
\ At a previous Snowmass Workshop it was argued \cite{Snowmass} that the
correct approach at hadron colliders should define jets based on nearness in
angle, or more specifically nearness in rapidity and azimuth. \ This idea is
the basis of the cone jet algorithm, which has been successfully employed to
test QCD at the Tevatron \cite{EKS},  
matching data to theory
at least at the 10\% level.
\ The second alternative is based on nearness in momentum space,
 and is typically called the
$k_{T}$ algorithm. \ It has been employed particularly at $e^{+}e^{-}$
colliders and more recently $ep$ machines. \ Suggestions \cite{ktcluster}\ that
the $k_{T}$ algorithm be used also at the Tevatron are also now yielding
results \cite{D0}. \ A characteristic feature of both of these algorithms is
that they identify and assign final state hadrons to unique jets
event-by-event. \ While this may seem like a perfectly natural procedure, it
is not without ambiguity, certainly if our goal is precision at the 1\% level.
\ The short-distance partons are, after all, QCD color nonsinglets while the
hadrons in the jets are color singlets. \ Thus the color singlet jets
constructed from the observed final states must arise from the correlated
evolution of color singlet \emph{combinations} of partons in the
short-distance scattered state, not from single partons. 
These effects are power-suppressed in $p_T$, but not necessarily negligible
for that reason \cite{Sope97}; see section \ref{sec:restopc} below.
\ Thus it is
essential that the jet analysis serve to suppress any sensitivity to this
inherent ambiguity, especially as it arises in the stochastic processes of
showering and hadronization. \ Of course, at lepton-hadron or hadron-hadron
colliders the color of the hard scattered parton leading to the jet could also
be balanced by color charge carried by remnants of the initial hadrons - the
spectator partons. \ For hadron-hadron colliders these spectators lead to the
very important and interesting question of the role of the underlying event,
which 
also factors from the hard-scattering process only to leading power. \ 
Since this contribution to the event does not decouple completely
from the hard scattering, and also contributes hadrons to the jets by any
measure of nearness, it must be understood if we are to reach our goal of
precision jet physics. \ Note that, while a perturbative analysis can say
something about the initial state and final state radiation contributions to
the underlying event, the contributions of the beam remnants are totally
outside of the scope of QCD\ perturbation theory 
at leading power.

\subsection{Underlying Event}

This important subject is treated in some detail in a separate contribution to
these Proceedings \cite{RDF}. \ The underlying event 
may be defined as everything
except the two most energetic jets in the final state, \textit{i.e}.,\ the
underlying event includes both initial and final state radiation and the
beam-beam remnants. \ The new results presented at the Workshop include a
detailed comparison of CDF\ 
data with results from three Monte
Carlo event simulators, ISAJET, HERWIG\ and PYTHIA. \ The analysis makes
insightful use of the concept of the ``transverse'' activity in the events,
where transverse here means transverse to the direction of the approximately
back-to-back energetic jets. \ If the leading jet defines $\phi=0$, the
transverse region corresponds to $60^{\circ}<\left|  \phi\right|  <120^{\circ
}$ in the same region of rapidity as the jets. \ On an event-by event basis,
the activity in the 2 transverse regions ($\phi>0$ and $\phi<0$) are ordered
as the ``transMAX'' and ``transMIN'' regions depending on the level of
activity, which are studied separately and as the sum and difference. \ The
transMAX\ region preferentially is correlated with the ``hard'' part of the
underlying event (initial and final state radiation plus extra jets), while
the transMIN\ region is correlated with the beam remnants, although the
separation is, of course, not precise. \ The ISAJET results, which assume
independent fragmentation, exhibit too many soft particles in the underlying
events and the wrong correlation with the $P_{T}$ of the leading jet. \ In
contrast, both HERWIG and PYTHIA include some level of color coherence effects
(angular ordering) in the parton showering. \ However, HERWIG, like ISAJET,
exhibits a $P_{T}$ spectrum for the beam remnant particles that is too steep
and thus does not generate enough remnant particles above $P_{T}=0.5$ GeV.
\ \ PYTHIA, which includes the effects of multiple parton scattering, seems to
come closest to agreement with the data, suggesting that such extra underlying
scattering processes are an important part of the correlation with the hard
scattering. \ 
That is, it appears that
hard scattering events are more central
collisions, and hence more likely to yield secondary parton scattering
processes. \ This conclusion, however, 
implies sensitivity to multiparton correlations.
\ Fully understanding the
underlying event is clearly
a demanding challenge.

\subsection{$k_{T}$ Algorithm}

Discussions of the hadron collider application of the $k_{T}$ algorithm
included a presentation 
 of the recent results from D\O\cite{D0}.
\ While these first results indicate 
reasonable agreement with
NLO\ theoretical expectations, the experimental results lie systematically
above the theory, especially for jet $E_{T}<150$ GeV. \ At least naively, this
can be ascribed to the expectation that the $k_{T}$ algorithm will tend to
``vacuum-up'' unassociated but nearby energy, which is not present in the
perturbative analysis. \ With our goal set at 1\% jet physics, it will be
important to test whether the $k_{T}$ algorithm can reach this precision.

\subsection{Cone Algorithm}

Detailed results concerning the efficacy of the cone algorithm were reported
at the Workshop, and are presented in a separate contribution to the these
Proceedings \cite{cone}. \ In the last year several previously unappreciated
features of the cone algorithm have come to light. \ Recall that the cone
algorithm identifies discrete jets event-by-event by requiring that ``found''
jets correspond to ``stable'' cones. \ The algorithm requires that the set of
hadrons (or calorimeter towers) that make up a jet not only lie within 
radius $R$ (in $\eta,\phi$ space) of the center of the cone ($\eta_{J}%
,\phi_{J}$) but also that the $E_{T}$ weighted centroid of the set coincide
with the center of the cone. \ In equations this looks like
\begin{eqnarray}
k &  \in J:\left(  \phi_{k}-\phi_{J}\right)  ^{2}+\left(  \eta_{k}-\eta
_{J}\right)  ^{2}\leq R^{2},\nonumber\\
\phi_{J} &  =\sum_{k\in J}\frac{E_{T,k}\phi_{k}}{E_{T,J}},\quad\eta_{J}%
=\sum_{k\in J}\frac{E_{T,k}\eta_{k}}{E_{T,J}},\label{cone}\\
E_{T,J} &  =\sum_{k\in J}E_{T,k}.\nonumber
\end{eqnarray}
It is 
this latter 
``stability'' constraint that results in the new issues.
\ When smearing is included, \textit{i.e}., when the energy of an underlying
short-distance parton is spread out in $\eta,\phi$, as presumably necessarily
happens due to showering and hadronization, stable solutions present at the
perturbative level disappear in the smeared final state. \ This occurs both in
a simple theoretical calculation of the smearing effect and in studies of more
realistic simulated final states. In the latter studies of Monte Carlo events,
sizeable amounts of nearby energy in the LEGO\ plot are simply not identified
with what is obviously an associated (\textit{i.e}., nearby) jet. \ While the
overall impact on jet rates is at only the 5 to 10\% level, this effect is
clearly relevant to 1\% jet physics. \ The study also confirms that a simple
``fix'' using a smaller cone size during the jet discovery phase, but not
during the jet construction phase, removes much of the problem when applied to
the Midpoint Algorithm recommended by the Run II Workshop \cite{workshops}.
\ On the other hand this fix is unlikely to help the Seedless
Algorithm \cite{workshops} address this issue. \ An interesting historical note
is that the traditional CDF Cone algorithm, JetClu
 \cite{JetClu}, has a
(largely undocumented) feature that seems to help with the difficulty
discussed here. \ JetClu
 (like the D\O \ cone algorithm) only looks for stable
cones in the vicinity of energetic calorimeter towers (or clusters of such
towers), the seed towers. \ Any towers inside an initial seeded cone remain
associated with that cone even as it migrates in $\eta,\phi$ towards a stable
location. (The cone algorithm is applied in an iterative process using the
$E_{T}$ weighted centroid for a given cone center to define the next cone
center until the two coincide.) \ The sticking of the towers to the initial
cone is labeled as ratcheting. \ Ratcheting is not simulated in any way in the
perturbative analysis.

\subsection{Jet Energy Flow}

In an effort to study a related but quite different approach to jet physics,
the Jet Working Group actively considered the jet energy flow (JEF) approach
and has made a separate contribution to these Proceedings \cite{JEF}. \ 
The JEF approach is an effort to explore, apply and extend a formalism introduced
and extensively developed
by Tkachov, in terms of observables based on caloriometric measurements,
which he termed ``C-continuous" \cite{Tkac96}.  In Ref.\ \cite{Tkac96}, the
algorithmic counting
of jets of definite energies and directions
is replaced by a class of continuous measurements of the flow of energy \cite{Tkac94}. 

\ As suggested above, a characteristic (and troubling) feature of traditional jet
algorithms is the event-by-event identification of individual hadrons with
discrete jets, and thus with a single (or color nonsinglet pair of) underlying
short-distance parton(s) in the perturbative calculation. \ We know this
connection cannot be made precise in detail, and we expect this mismatch to
enhance the sensitivity to the effects of showering and hadronization (where
the extra color correlations must appear). \ The issues raised above
concerning the cone algorithm are presumably of just this nature. \ Thus it is
appealing to consider other procedures for mapping the hadronic final state
onto the underlying short-distance partonic state. \ The JEF approach is just
such a procedure. It accepts the reality that the hadronic final state
represents the collective radiation from several out-flowing color charges,
\textit{i.e}., the underlying short-distance partons. \ No attempt is made to
associate individual hadrons with unique jets, \textit{i.e}., with unique
underlying partons, on an event-by-event basis. \ Yet the energy flow pattern
of an event still provides a footprint of the underlying partons, from which
much of the same information provided by the standard jet algorithms can be
extracted. \ The basic idea is to generate a jet energy flow distribution to
describe each event instead of a discrete jet of jets. \ This is accomplished
by convoluting the observed energy distribution with a simple smearing
function as suggested in the figure, where the smearing function in this case
is just a constant inside the circle (cone) and zero outside. \
\begin{figure}
[ptb]
\begin{center}
\vbox {\vskip 1 cm \hbox{\hskip 1 cm \includegraphics[
natheight=5.955100in,
natwidth=7.726200in,
height=3.9894in,
width=5.1716in
]%
{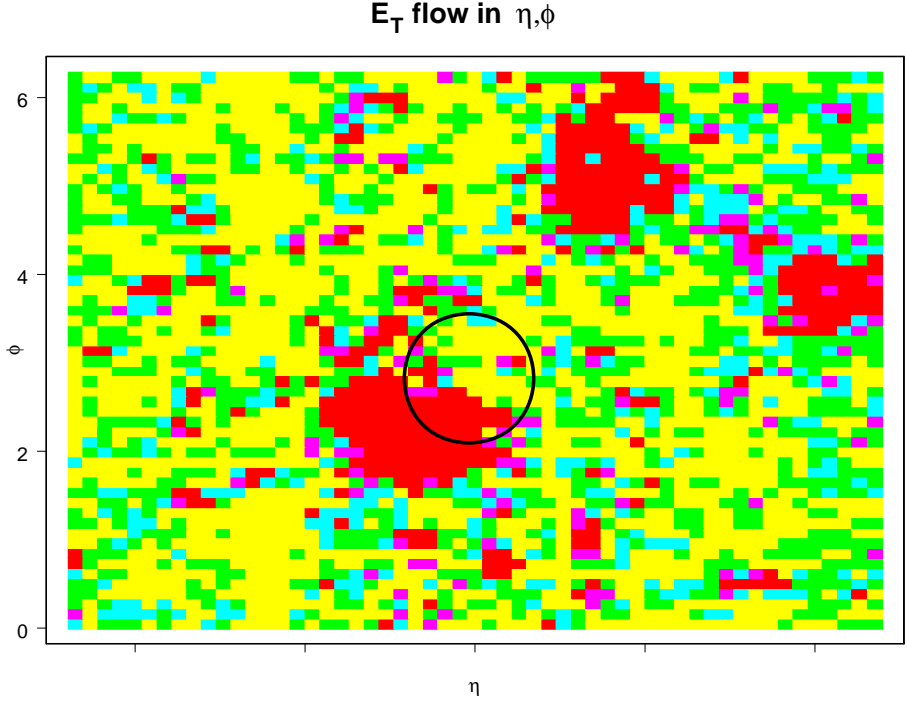}}}%
\caption{Plot of energy flow in $\eta,\phi$ indicating the region averaged
over to determine the JEF\ value at the center of the circle.
In this figure, red (dark) pixels indicate high energy.
The event was created with PYTHIA.}%
\label{fig:JEF}%
\end{center}
\end{figure}
In equations, we start with the underlying 4-vector distribution for the event
(the color coded pixels in the figure) defined by
\begin{equation}
P_{\mu}\left(  \widehat{P}\right)  =\sum\limits_{i=1}^{N}p_{\mu}^{i}%
\delta\left(  \widehat{P}-\widehat{P^{i}}\right)  ,\label{4-vector}%
\label{pmu}
\end{equation}
where the directional unit vector is defined by $\widehat{P}\left(  m\right)
=\overrightarrow{P}/\left|  \overrightarrow{P}\right|  $ with the
2-dimensional angular variable defined as $m=\left(  \eta,\phi\right)  $
(typical of hadron colliders, where $\eta$ is the pseudorapidity, $\eta
=\ln\left(  \cot\theta/2\right)  $) or $m=\left(  \theta,\phi\right)  $
(typical of lepton colliders).  
Eq.\ (\ref{pmu}) is the building block from which all weights and
correlators of energy flow may be constructed \cite{Tkac96}.
For example, it 
defines the underlying energy
flow via $E\left(  m\right)  =P_{0}\left(  m\right)  $ or (for the case of
hadronic colliders) the more familiar underlying transverse energy flow is
defined by the composite quantity $E_{T}\left(  m\right)  =\sqrt{P_{x}%
^{2}\left(  m\right)  +P_{y}^{2}\left(  m\right)  }$ or $E_{T}\left(
m\right)  =E\left(  m\right)  \times\sin\theta\left(  m\right)  $. \ Clearly
many quantities can be constructed from the 4-momentum distribution of Eq.
\ref{4-vector}, including the usual cone jet algorithm. \ We define the jet
energy flow (JEF) via a smearing or averaging function $A$\ as%
\begin{equation}
J_{\mu}\left(  m\right)  \equiv\int dm^{\prime}\ P_{\mu}\left(  m^{\prime
}\right)  \ \times\ A\left(  m^{\prime}-m\right)  ,\label{eq:JEF}%
\end{equation}
where $A$ is normalized as%
\begin{equation}
\int dm\ A\left(  m\right)  =1.\label{norm}%
\end{equation}
A simple (but not unique) form for the averaging function in\ terms of the
general 2-tuple of angular variables $m=\left(  \alpha,\beta\right)  $, which
provides a direct comparison with the jet cone algorithm, is%
\begin{equation}
A\left(  m\right)  =A\left(  \alpha,\beta\right)  =\frac{\Theta\left(
R-r\left(  \alpha,\beta\right)  \right)  }{\pi R^{2}}=\frac{\Theta\left(
R-\sqrt{\alpha^{2}+\beta^{2}}\right)  }{\pi R^{2}},\label{averaging}%
\end{equation}
where $R$ is the cone size and $r\left(  \alpha,\beta\right)  $ is the
distance measure in the space defined by $\left(  \alpha,\beta\right)  $.
\ The specific example of the jet transverse energy flow (transverse JEF)%
\begin{equation}
J_{T}\left(  m\right)  =\int dm^{\prime}\ E_{T}\left(  m^{\prime}\right)
\ \times\ A\left(  m^{\prime}-m\right)  =\int dm^{\prime}\ \sqrt{P_{x}%
^{2}\left(  m^{\prime}\right)  +P_{y}^{2}\left(  m^{\prime}\right)  }%
\ \times\ A\left(  m^{\prime}-m\right)  \label{JetET}%
\end{equation}
for the case of $R=0.7$ is suggested in the figure. \ Note that the transverse
JEF is smeared on a scale $R$ compared to the underlying $E_{T}$ distribution.
\ For comparison, the Snowmass cone jet algorithm \cite{Snowmass,EKS} identifies
jets at a discrete set of values of locations $m_{j}$ defined by the $E_{T}$
weighted cone ``stability'' constraint. \ These stable cone locations $m_{j}$
are the solutions of the equation%
\begin{equation}
\int dm^{\prime}\ E_{T}\left(  m^{\prime}\right)  \ \times\ \left(  m^{\prime
}-m_{j}\right)  \ \times\ A\left(  m^{\prime}-m_{j}\right)
=0.\label{stablecone}%
\end{equation}
The corresponding cone jet $E_{T}$ values are found by evaluating Eq.\
(\ref{JetET}) (times $\pi R^{2}$) at the jet positions, $E_{T,j}=\pi R^{2}\times
J_{T}\left(  m_{j}\right)  $. \ 

A primary strength of the JEF approach is that, in contrast with the usual
algorithmic approach
to jet identification,
the JEF formalism 
generates a smooth distribution, $J_\mu(m)$, event-by-event,
and, in that sense, the JEF formalism is more
analytic. For example, in the application of the cone algorithm the goal of
identifying unique jets leads to the ``stability'' constraint as noted above.
\ The non-analytic character of the jet cone algorithm arises from the need to
solve Eq.\ (\ref{stablecone}) and then work with only the discrete set of
solutions, \textit{i.e}.,\ the jets in an event. \ Only limited (and typically
complicated) regions of the multiparticle phase space contribute to a jet.
\ No such constraint arises in the JEF\ analysis. This distinction has several
important consequences, including the expectation that the more inclusive and
analytic calculations characteristic of a JEF analysis are more amenable to
resummation techniques
and power corrections analysis \cite{Korc99}
 in perturbative calculations; that, since the required
multiparticle phase integrations are largely unconstrained, \textit{i.e.,}
more analytic, they are easier (and faster) to implement; that since the
analysis does not identify jets event-by-event, the analysis of the
experimental data from an individual event should proceed more quickly; and
that the signal to background optimization can now include the JEF\ parameters
(and distributions). \ 
As emphasized in \cite{Tkac96},
the representation of the final state provided by the
JEF, or more generally in terms of C-continuous observables,
 is expected to be a more reliable footprint in the sense of exhibiting a
reduced sensitivity to the showering and hadronization processes. \ The
challenge in the JEF type analysis is to define quantifiable observables.
\ \ Several simple examples are provided in the separate contribution to
these Proceedings.

\subsection{Conclusions}

To make the best possible application of the increased precision of our
understanding of QCD outlined elsewhere in this Summary, it is essential the
our understanding of jets and how to identify them also improve. \ Several
relevant issues were addressed at the Workshop and are discussed briefly in
this summary and elsewhere in the Proceedings. \ The overall conclusion is
that 1\% jet physics is an important and possible goal but there is much work
still to be done.

%%%%%%%%%%%%%%%%%%%%%%%%%%%%%%%%%%%%%%%%%%%%%%%%%%%%%%%%%%%%%%%%%%%%%%%%%%
\section{Resummation}
\label{sec:resum} 

{\it This section was prepared with the participation of
Csaba Balazs, Anna Kulesza,
Nikolaos Kidonakis, 
Pavel Nadolsky and George Sterman.} 

%%%%%%%%%%%%%%%%%%%%%%%%%%%%%%%%%%%%%%%%%%%%%%%%%%%%%%%%%%%%%%%%%%%%%%%%%%
\subsection{Threshold Resummation} \label{sec:resumi}

The goal of threshold resummation is to resum potentially large logarithms
that appear in the cross section for production of a given
high-mass system, $F$, near threshold, to all orders in the strong coupling 
$\alpha_s$.
Choices for $F$ include Higgs and weak vector bosons, heavy
quarks and high-$p_T$ jet pairs, with fixed or integrated rapidities.
The logarithms in question arise from the incomplete 
cancellation of infrared divergences between virtual and real graphs near 
threshold in the partonic
system $a+b\rightarrow F+X$, where there is restricted phase space for real 
gluon emission.

Threshold resummation is derived from the factorization properties  
of the hadronic cross 
section~\cite{Berger:1995xz,Berger:1996ad,Kidonakis:1996aq,Kidonakis:1997gm,Kidonakis:1998bk,Kidonakis:1998nf,Laenen:1998qw}. For
example, the heavy quark cross section can be written as
\begin{equation}
\sigma_{h_Ah_B\rightarrow QX}=\int dx_a \, dx_b \;
f_{a/h_A}(x_a,\mu_F^2)
\; f_{b/h_B}(x_b,\mu_F^2) \;
{\hat \sigma}_{ab\rightarrow QX} \left(s_4,t_i,m^2,\mu_F^2,
\alpha_s(\mu_R^2)\right)  ,
\label{factcs}
\end{equation}
For the production of heavy quark $Q$, we define $t_i=(x_ip_i-p_Q)^2-m_Q^2$,
with $i=a,b$.  As usual, the non-perturbative parton distributions 
$f_{i/h_I}$ describe
the long-distance physics that is factorized from the hard scattering,
and ${\hat \sigma}_{ab\rightarrow QX}$ is the perturbative
partonic cross section, which still displays sensitivity to 
gluon dynamics that is perturbative, but soft compared to $m_Q$. 
This sensitivity manifests itself in plus
distributions of the type 
$[(\ln^m(s_4/m^2))/s_4]_+$, $m \le 2n-1$ at $n$th order in
$\alpha_s$.  The variable $s_4 \equiv s+t_a+t_b$, and $s_4$ vanishes 
at partonic threshold. 

A further re-factorization separates the non-collinear
soft gluons, described by a soft gluon function $S$, which is a matrix
in color space, from the hard scattering $H$. From the
renormalization properties of these functions one may obtain evolution   
equations that lead to the resummed cross 
section~\cite{Kidonakis:1996aq,Kidonakis:1997gm,Kidonakis:1998bk,Kidonakis:1998nf,Laenen:1998qw}.
Generally, the resummed cross sections are specified in a
Mellin or Laplace moment space, with the latter defined by 
$\hat{\sigma}(N)=\int (ds_4/s) \,
e^{-Ns_4/s} {\hat\sigma}(s_4)$, with $N$ the moment variable.
Note that under moments,
$[(\ln^{2n-1}(s_4/m^2))/s_4]_+ \rightarrow \ln^{2n}N$, and our aim
becomes to resum logarithms of $N$.

The resummed cross section at next-to-leading logarithmic (NLL)
and higher accuracy can be written as~\cite{Kidonakis:2001mc}:
\begin{eqnarray}
{\hat{\sigma}}_{ab \rightarrow QX}(N) &=&
\exp\left\{ \int_{Q^2/N^2}^{Q^2}{dm^2\over m^2} 
\left[\, A_a(\alpha_s(m))+A_b(\alpha_s(m))\, \right]\, \ln\left({Nm\over 
Q}\right)\,
\right\} \nonumber\\
&\ & \hspace{-10mm} \times {\rm Tr} \left \{H
\exp \left[\int_m^{m/N} \frac{d\mu}{\mu} \, \Gamma_S^{\dagger}\right]
{\tilde S} \left(\alpha_s(m^2/N^2) \right)
\exp \left[\int_m^{m/N} \frac{d\mu}{\mu} \, \Gamma_S \right] \right\}\, ,
\end{eqnarray}
where the first factor gives the double-logarithmic $N$-dependence from the 
incoming partons, in terms of anomalous dimensions 
$A_a(\alpha_s)=(\alpha_s/\pi)C_a+\dots$ with
$C_q=C_F$ and $C_g=C_A$,
while $\Gamma_S$ is an anomalous dimension matrix
of soft gluons, which controls single logarithms
associated with coherent interjet radiation~\cite{Kidonakis:1996aq,Kidonakis:1997gm,Kidonakis:1998bk}.

Threshold resummation was first derived for the Drell-Yan cross 
section~\cite{GS87,CT89},
and it has been applied to a large number of processes in
the last five years. Resummed cross sections and their finite-order 
(NNLO and higher) expansions have now been derived for
heavy quark
hadroproduction~\cite{Kidonakis:1997gm,ebhc2,BCMN98,KV99,Kidonakis:2000ui,Kidonakis:2001nj}  and
electroproduction~\cite{LM99,Iv01},  direct photon production~\cite{Laenen:1998qw,CMNOV99,KO00,SV01}, 
$W$ + jet production~\cite{KD00},
and dijet~\cite{Kidonakis:1998bk,Kidonakis:1998nf} and single-jet~\cite{KO01}
production. For a review see Ref.~\cite{NKrev00}.
It has been found that the threshold corrections are more or less
sizable, depending on the specific process 
and kinematics, and that the factorization
scale dependence of the cross sections for these processes is significantly 
reduced relative to NLO. However, ambiguities remain in the cross section, 
due 
to contributions from subleading logarithms~\cite{ebhc2,Kidonakis:2000ui} and due
to uncertainties from the kinematics (single-particle-inclusive
versus pair-inclusive)~\cite{Kidonakis:2001nj}.

%%%%%%%%%%%%%%%%%%%%%%%%%%%%%%%%%%%%%%%%%%%%%%%%%%%%%%%%%%%%%%%%%%%%%%%%%%
\subsection{${k_T}$ Resummation} \label{sec:resumii}

The soft-gluon resummation formalism for the inclusive production of
colorless massive states~\cite{Collins:1985kg} may be the most mature,
best understood, and best tested formulation among the resummation
techniques.  Its phenomenological advantage over threshold resummation is that
the effects of resummation are readily observable in the distribution
of the transverse momenta. Indeed, it is not possible, as in the case
of threshold resummation, to define a finite differential cross section
without resummation.

Recently both theoretical refinement and 
phenomenological applications of the original formalism have received
considerable attention. On the theory side, the improvement of its
non-perturbative sector~\cite{Qiu:2001hf}, the introduction of 
resummation schemes~\cite{Catani:2001vq}, and the development of 
combined threshold and transverse momentum ($Q_T$) 
resummation~\cite{Laenen:2000ht} are  important milestones. The
formalism is successfully applied to $W^\pm$, $Z^0$ boson, 
diphoton~\cite{Balazs:1997xd}, heavy quark pair~\cite{ebrm94}, and direct 
photon production~\cite{Kidonakis:2000hq} at the Tevatron, to 
DIS~\cite{Nadolsky:2000kb}, and to Higgs production at the 
LHC~\cite{Balazs:2000wv}. The latter application plays a critical role in
the search for a light Higgs boson. At Snowmass the results of the
full NLL calculation of the Higgs $Q_T$ distribution were 
presented~\cite{BalazsInThisProc}.

In recent papers~\cite{deFlorian}, Catani, de Florian and Grazzini observed
that NNLO soft terms in resummed small-\( p_{T} \) cross sections can be
obtained by combining coefficients of lower orders and two-loop splitting
functions. Their result substantially simplifies calculation of higher-order
terms for resummed cross sections in a variety of processes, including 
Higgs boson production at the LHC. When included in the Higgs boson cross
section, the soft NNLO terms significantly alter the shape of the small-\(
p_{T}\)
distribution~\cite{BalazsMoriond}, which suggests the importance of even
higher-order
corrections (and of more work to achieve adequate understanding of this
observable).

In the formalism of Ref.\ \cite{Collins:1985kg}, resummed cross sections
receive relatively unknown non-perturbative 
``power" contributions in the impact parameter $b$ from the region
of large $b$. Qiu and Zhang~\cite{QiuZhangprl,Qiu:2001hf} suggested that
the need for such contributions can be reduced by extrapolation of leading 
logarithmic
terms from the perturbative region, combined with the lowest-order power
corrections.
They found good agreement of their approach with data on vector boson
production. As an alternative to the numerical Fourier transform from the
impact parameter space, one may try to estimate sums of leading logarithmic
series directly in transverse momentum space. Kulezsa and Stirling
\cite{Kulesza2001} proposed a method for the systematic summation of leading
logarithmic towers in \( p_{T} \)-space while preserving some advantages
(\emph{e.g.}, transverse momentum conservation) of the \( b \)-space
formalism.
They find that the Tevatron \( W \) and \( Z \) data can be described
reasonably
well by the lowest four towers of logarithms, except at 
\( p_{T}\leq\) 2--3~GeV, where non-perturbative effects become important.
The method of joint resummation \cite{Laenen:2000ht} provides some of the 
advantages of both these
approaches, with a cross section that can be defined consistently
in perturbation theory, and a consistent picture of
nonperturbative corrections for both $k_T$ and threshold
resummations.

\subsection{Resummation for Final States: Power Corrections}
\label{sec:restopc}

Jet properties and event shapes
are another successful application for resummation techniques.
Technically similar to threshold resummation, they
share with $k_T$ resummation a direct relation
to observables.

Consider the resummed cross section for
$\rm e^+e^-\rightarrow Z^*(Q)\rightarrow F$, in the
elastic limit, where the final state $F$ consists of
two jets with invariant masses much less than $Q$.
In terms of the familiar ``thrust" variable, we
take the limit $1-T \sim (m_1^2+m_2^2)/Q^2 \rightarrow 0\, .$
In this case, the
light-like relative velocities between the jets
imply factorization, evolution and, eventually
a cross section resummed for the singularities in $t\equiv 1-T$,
\begin{eqnarray}
{1\over \sigma_{\rm tot}}\, {d\sigma_{\rm PT}(1-T)\over dT}
\sim
     {d\over dT}\;
\exp\;
\left[- \int_{t}^1 {dy\over y} \int_{y^2Q^2}^{y Q^2}
{d m^2\over m^2}\; A(\as(m))\right]\, ,
\label{Tresum}
\end{eqnarray}
shown here to leading logarithm.  The function $A$ 
is exactly the same function of the coupling encountered
in threshold resummation.

In the resummed cross section of (\ref{Tresum}),
we encounter 
the ``Landau" pole of the perturbative
running coupling in the limit $y\rightarrow 0$.
In the analysis of power
corrections, we make a virtue of the necessity of dealing
with the Landau pole.  Roughly speaking,
we discover that, unaided,
perturbation theory is not self-consistent, and
we search for a minimal class of nonperturbative
corrections to restore self-consistency \cite{irr}.  This
approach has had considerable phenomenological successes,
and is
beginning to shed light on the
relationship between perturbative and nonperturbative
dynamics in QCD. 

More insight can be gained by expressing the exponent
in more detail.  The following form is equivalent to
(\ref{Tresum}) at leading logarithm, but actually
incorporates all logarithmic behavior in the eikonal
approximation,
\begin{eqnarray}
   \int_0^1 dy{{\rm e}^{y/t}-1 \over y} \int_{y^2Q^2}^{y Q^2}
{d m^2\over m^2}\; A_q(\as(m)) \sim  {{2\over tQ}\int_0^{\mu_0} dm\;
A_q(\as(m))} + \dots
\, ,
\label{Texpand}
\end{eqnarray}
where on the right we have exhibited the leading behavior in $Q$ of
the integral for low values of the momentum scale $m$,
with $\mu_0$ a fixed cutoff $>\Lambda_{\rm QCD}$.
In these terms, it is natural to think of the
$1/Q$ term as the dominant
power correction implied by
the ambiguity of the perturbative
series.  In addition, because the function $A$
appears often (recall threshold resummation), 
it is tempting to think of 
the correction as in some sense universal.
\begin{figure}[h]
%\vbox{\vskip 1 true in}
\hbox{\hskip 3.0 cm
\includegraphics[width=20pc]{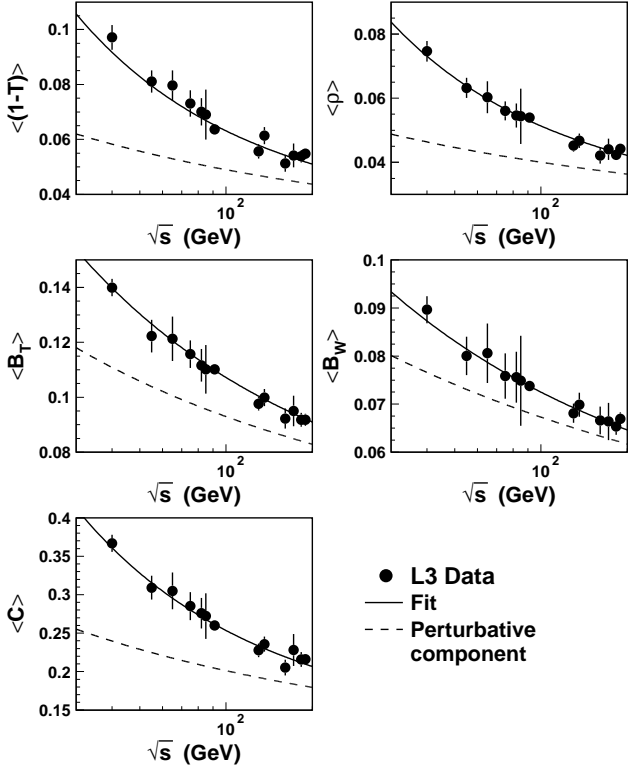}}
\caption{L3 data for first moments of various
event shapes. The dashed line is perturbation
theory, the solid line a fit based on
Eq.\ ({\protect \ref{shift}})}.
\label{l3}
\end{figure}

These ideas have been tested most extensively in
electron-positron annihilation, making use of the
  energy lever arm of the LEPI and II programs.
Resummations similar to that for thrust can be carried out for a
variety of event shape functions measured at LEP and
elsewhere, which measure various kinematic properties of jets,
and which are labelled by a small alphabet soup of
letters, which we denote by $e=1-T,B_T,C \dots$.
All of the $e$'s are defined to vanish in
the elastic limit.

As a minimal approach, we assume that we only have to
worry about the first term in the expansion (\ref{Texpand}),
In general, the overall perturbative
coefficients of the integral over the running coupling
differ from event shape to event shape.
Taking this into account, we define \cite{Dok98},
\begin{equation}
\lambda_e \sim C^{(e)}_{\rm PT}{\cal M}_e\int_0^{\mu_0} dk\, \as(k)\, ,
\label{lambda}
\end{equation}
where, adopting one popular notation, we replace the
function $A(\alpha_s)$ by the combination ${\cal M}_e\as$.
The constant ${\cal M}_e$ incorporates certain higher-order
effects that depend, in part, on the event shape in question.
In this formalism we identify a
``universal" nonperturbative parameter
as the integral of the running coupling from zero to
some fixed scale $\mu_0$, which again has the
interpretation of a factorization scale.

Adding such a term to a suitably-defined perturbative exponent produces
a simple shift of the cross section,
\begin{eqnarray}
{d\sigma(e) \over de} =
{d\sigma_{\rm PT}(e-\lambda_e/Q) \over de} +{\cal O}\left({1\over 
e^2Q^2}\right)\,
.
\label{shift}
\end{eqnarray}
This result gives a number of consequences, for
instance, that the true first moments of event shape $e$
are related to the perturbative predictions
for the first moments by $\langle e \rangle = \langle e
\rangle_{\rm PT} + {\lambda_e/Q}$,
with similar predictions for second moments.

The phenomenology of various event shapes
  \cite{L300}  shows quite good agreement
with this picture for the first moments, as shown
in Fig.\ \ref{l3}.  At the same time, there are
unexpectedly large $1/Q^2$
corrections in second moments, such
as $\langle(1- T)^2(Q)\rangle$.
First moments, however, are relatively blunt
instruments to study the perturbative-nonperturbative 
interface, and there are interesting alternative
explanations of the data \cite{delphi490}.

A natural extension of the shift (\ref{shift}),
to take into account soft radiation from dijets
beyond the pure $1/Q$ correction, is to
generalize the shift to a convolution \cite{Korc99},
\begin{eqnarray}
{d\sigma \over de} =
\int_0^{eQ} d\epsilon\; f_e(\epsilon)\; {d\sigma_{\rm PT}(e-\epsilon/Q) \over
de}  +{\cal O}\left({1\over eQ^2}\right)\, ,
\label{shapeconvol}
\end{eqnarray}
where the nonperturbative
``shape function" $f_e$ is independent of $Q$.
It is therefore sufficient to fit $f$ at $Q=m_Z$
to derive predictions for all $Q$.
A justification for (\ref{shapeconvol})
may be found in an  effective theory
for soft gluon radiation by $3\otimes 3^*$ sources,
which represent a  quark pair, and are
given by products of ordered exponentials
in the directions of the jets' momenta.
The  phenomenology of shape functions has
been discussed in
\cite{Gar01} and in \cite{Taf01},
while in \cite{Bel01},
it was shown that the double distribution
determined phenomenologically in \cite{Taf01},
follows from rather general considerations.
In considerations such as these, resummation
opens the path from perturbative to
nonperturbative QCD.

The extension of these methods to hadronic scattering
has begun relatively recently \cite{banf01,berg01}.   This effort should
enable us to use energy flow analysis
in jet and other hard-scattering events to
make hadronic final states more accessible to
both perturbative and nonperturbative QCD.
In studying differential hadronic final states,
we may also hope to promote techniques introduced
for threshold resummation to a greater phenomenological 
relevance.  To make use of these ideas will require 
an approach to hard-scattering events closer
to the philosophy of ``jet energy flow", described
in Section \ref{sec:jet} above.  
The textures of events contain a wealth of
information that may be lost if they
are analyzed for single purposes and then
forever discarded.  This suggests the daunting challenge of
somehow ``archiving" data for a future in which
new ideas will arise, and be answered by
asking new questions.  

%%%%%%%%%%%%%%%%%%%%%%%%%%%%%%%%%%%%%%%%%%%%%%%%%%%%%%%%%%%%%%%%%%%%%%%%%%

\section{Lattice QCD}
\label{sec:lqcd}

{\it This Section was written by Paul Mackenzie, who thanks
Peter Lepage and Andreas Kronfeld for their input.}

QCD is the theory of all of strong interaction physics, 
but the traditional perturbative methods of field theory
are only suitable for investigating strong interactions
at short distances, where asymptotic freedom causes
 the effective coupling constant to be small.
Lattice gauge theory provides a way of understanding
the effects of strong interactions
at all distance scales, including the long distance effects
that are the hallmark of strong interactions and for which
traditional methods fail.
Accurate lattice calculations are necessary for showing that we
understand QCD at long as well as at short distances.
Lattice calculations are required for extracting non-QCD
standard model effects  such as 
Cabibbo-Kobayashi-Maskawa (CKM) matrix elements
from experiment.
They also provide  prototypes for investigations of nonperturbative
effects in strongly coupled
 beyond-the-standard-model theories, which have been 
little studied so far because of the incompleteness of
field theoretic methods.

Lattice QCD has gone through a revolution in the last five to ten years.
Ten years ago, there were very few phenomenological lattice results 
that were solid enough to be of much interest to particle physicists outside
the lattice community.
Today, there are many such results in the quenched approximation
(omitting the effects of light quark loops), and decent unquenched 
calculations are becoming more and more common.
Solid  results are appearing for some of
the nonperturbative quantities of greatest importance to standard
model phenomenology, such the amplitudes for determining 
CKM matrix elements.
In comparison with perturbative QCD calculations,
solid lattice calculations were slow to develop.
After the invention of QCD,
the methods of perturbation theory were applied to short distance
strong interactions physics and produced quantitative results
immediately, in
the parton distribution functions of deep inelastic scattering for example.
The application of QCD to nonperturbative long distance phenomena, by 
contrast, required the development of new methods and took much longer.
Wilson proposed formulating QCD on a lattice soon
after the discovery of QCD in 1973, with the aim
of making a wider array of calculational methods applicable.
Almost ten years passed before the establishment of the current 
calculational paradigm of large-scale Monte Carlo simulations of the
QCD path integral, and more than another ten years after that before reasonably
accurate phenomenological results became common.

\subsection{The Path to High Precision}

 The accuracies so far achieved for  important phenomenological
 calculations
are far below what will be required by experiment.
For example, $B$ factories and collider experiments
will determine the experimental amplitudes for $B\overline{B}$
oscillations in $B_d$ and $B_s$ mesons to an accuracy of 1\% 
or better.  
Determining the controlling CKM matrix element requires lattice
calculations of the associated hadronic amplitudes to the same accuracy,
far beyond current lattice standards of perhaps 10\%.
For all the progress that has been made in making lattice
calculations more precise, much more is required.

The ever-increasing power of computers,
from around one floating point operation/second/dollar (flops/\$) 
 on the Vax 11/780s on which the first
Monte Carlo spectrum calculations were done,
to around 1,000,000 flops/\$ on one of today's Pentium 4 servers,
has been an essential element of the improving accuracy of 
lattice calculations.
However, the accuracy $\epsilon$ improves  slowly with CPU power, typically 
as $\epsilon \sim {\rm CPU\ power}^{-1/p}$ where $p$ is somewhere between
six and nine.
Investigation of algorithms that minimize this scaling power
is an important and ongoing avenue of lattice research \cite{pea01}.
 Even an order of magnitude increase in  computing
power  improves accuracy by only tens of per cent, so more than
brute force computing improvement alone 
is necessary to bring us to the required accuracy.
Increasingly accurate approximations to the normalization factors
$Z(\mu a,\alpha_s(\mu))$ and the higher dimension correction operators
$K_n(\mu,a,\alpha_s(\mu))$ in Eqn.~(\ref{P5_lattice})
%%%%%%%%%%% Equation number hard-wired.
have also been required for higher precision lattice calculations,
and continue to be required for further progress.

There are three main families of lattice actions in current use
for light fermions.
Wilson fermions can be formulated for any number of quark flavors,
at the expense of adding an ${\cal O}(a)$ discretization error
to the quark action.
A nonperturbative renormalization program for Wilson fermions has been
carried out through ${\cal O}(a)$ corrections~\cite{lue97}.
Staggered and naive fermions have only even dimensional correction
operators and are relatively quick to compute in unquenched simulations,
but are naturally formulated with four or sixteen degenerate flavors
in the fermion determinant \cite{tou01}.
Calculations with physical numbers of flavors must be done
by taking a root of the determinant, eliminating the possibility of
formulating the theory in terms of a local action and thus requiring
more careful attention to the locality of the resulting theory. The use of domain 
wall/overlap fermions has made it possible to
formulate lattice theories with exact chiral symmetry, and without
fermion doubling, which was the most significant technical challenge in
lattice QCD left unsolved by Wilson \cite{neu01}. This allows a clean
definition of the bare quark mass, avoidance of chirality forbidden
mixings in lattice operators, and other advances. They have only even
dimension correction operators, but are more expensive to simulate since
they introduce and extra space-time dimension or a large number of
additional fermion fields. All three families of actions
are currently under active investigation.

As with all applications of QCD, 
high precision lattice calculations will require next-to-next leading
order perturbation theory (or its nonperturbative lattice equivalent)
for the normalizations $Z(\mu a,\alpha_s(\mu))$ in Eqn.~(\ref{P5_lattice}).
%%%%%%%%%%% Equation number hard-wired.
Although lattice perturbation theory
 expressed in terms of the bare coupling
constant often converges poorly, when formulated properly
it converges at roughly the same rate as
 dimensionally regulated perturbation theory~\cite{lepxx}.
Automated perturbation theory techniques make possible the
efficient handling of the complicated actions of
lattice gauge theory \cite{lue}.
Furthermore, more short distance tools are available
on the lattice than in dimensional regularization.
Nonperturbative short distance calculations may be used to
test and extend conventional lattice perturbation theory calculations.
Short distance lattice calculations may be formulated entirely
nonperturbatively, so that only the dimensionally regularized part
of a lattice--$\overline{MS}$ matching is done perturbatively \cite{lue97}.

\subsection{Applications of Lattice Gauge Theory}

Lattice QCD calculations serve to test lattice methods (and QCD, too, if you
prefer), to extract standard model information (such as quark masses 
and CKM matrix elements) from data, and to serve as prototypes for
investigations of strongly coupled beyond-the-standard-model theories.
For recent reviews of applications of lattice gauge theory, see  \cite{lat01}.
Some of the most interesting and important applications of lattice QCD 
are also among the most solid.
 Determinations of the strong coupling constant, the quark masses,
and the CKM matrix elements can all be made 
using lattice calculations having a single, hadronically stable meson
on the lattice at a time.  For example,
most constraints on the $\rho-\eta$ plane
%, which parameterizes
%$CP$ violation in the standard model, as shown in
%Fig.~\ref{fig:rhoeta}, 
will ultimately be governed by such lattice QCD
calculations.
These can be considered golden calculations of lattice QCD.
Stable particles are easier to analyze than unstable particles.
Stable mesons are small (thus, having smaller finite volume errors)
and light (thus, having smaller statistical errors).
The heavy quark masses and strong coupling constant are particularly
easy to obtain since they can be obtained from heavy quarkonia, where
nonrelativistic methods are also available to help analyze corrections.
Good prototype calculations exist for all the quantities in
this category, and no roadblocks are known to ever increasing accuracy.
Many stable meson masses are known
experimentally with high precision that can test
lattice mass predictions, as well as provide determinations of quark
masses and the strong coupling constant.
Lattice calculations can use data from
BaBar, Belle, and the Tevatron to provide accurate determinations
of CKM parameters.
These data do not provide any high precision tests of lattice amplitudes,
however.
Cleo-c will provide new lattice-based determinations of CKM matrix elements
in the charm sector, and will also provide some much more accurate tests
of lattice amplitudes, for example
using the ratio of the $D$ meson decay constant
and the semileptonic amplitude $D\rightarrow \pi l \nu$.

All of nonperturbative QCD is ultimately approachable through lattice
calculations, and not just these simplest quantities.
Recent progress has been made in formulating the lattice amplitudes
for kaons decaying into
multiple pions, required for lattice calculations
of the $CP$ violation parameter $\epsilon'/\epsilon$ \cite{has01}.
More work is required to formulate the equally important amplitudes when
the decaying particle is heavy, such as a $B$ meson.
QCD at finite temperature and density is an ongoing area of research
\cite{han01}.

Lattice QCD calculations are exciting not only because of their direct
usefulness, but also as prototypes of general nonperturbative field theory
calculations. QCD exhibits many of the properties expected in other
strongly coupled theories: confinement, chiral symmetry breaking,
instantons, etc. Since many solutions of the gauge hierarchy problem, both
supersymmetric and nonsupersymmetric, rely on a strongly coupled sector of
a new gauge interaction, this is an area of phenomenological as well as
theoretical importance. As noted above, the formulation of theories with chiral fermions
was an outstanding unresolved issue in lattice field theory.  In the
last few years, the problem of formulating a vector-like theory with an
exact chiral symmetry has been solved.  It has also been shown how to
formulate N=1 supersymmetryic Yang-Mills theories (in which the fermions
are real) \cite{mon01}.  Enormous progress has been made in the last few
years in the development of nonvector-like chiral gauge theories
\cite{gol00, neu01}.  Active research is underway investigating the
remaining technical and algorithmic challenges to performing Monte Carlo
simulations for theoires with complex actions.

\subsection{Computing Facilities for Lattice QCD}

Large scale computing facilities are a crucial component
of accurate lattice calculations.
The scale of the required facilities, although small on the
scale of high energy physics experiments, is larger
than usual for theoretical physics research.
Multiteraflop projects have been established in Europe and Japan
for this purpose.
To meet these challenges in the US, the DoE has established
a coordinated planning process for US lattice computing needs.  
It is led
by an executive committee consisting of Bob Sugar (chair),
Norman Christ, Mike Creutz, Paul Mackenzie, John Negele, Claudio
Rebbi, Steve Sharpe, and Chip Watson \cite{sug00}.
The two most important avenues for hardware are being investigated,
large clusters of fast, cheap commodity hardware such as
Pentium servers, and more tightly integrated purpose built 
equipment.
The DoE's Scientific Discovery through Advanced Computation
(SciDAC) program has funded a software development program and
 medium-sized prototype hardware,
which is now being constructed.
Planning is underway for terascale hardware in '03/'04.

\subsection{Summary}
Lattice calculations have become an essential quantitative tool
for strong interaction physics.
For some of the most important lattice quantities
in standard model phenomenology, increasingly precise
results are immediately foreseeable.
These calculations require computational resources that are larger
than has been customary for theoretical physics, but
are still tiny compared to the the experiments that require them.

%%%%%%%%%%%%%%%%%%%%%%%%%%%%%%%%%%%%%%%%%%%%%%%%%%%%%%%%%%%%%%%%%%%%%%%%%%

\section{From Small $x$ to Nuclei}
\label{sec:smallxnuc}

{\it This section written by Ina Sarcevic, Jamal
Jalilian-Marian and George Sterman.}

Unlike electrons or quarks, hadrons have structure, an inside and an outside, 
stabilized by ongoing interactions. Their factorized, partonic description, 
however, is in terms of noninteracting
states. The interactions that bind the hadron are expressed only in the 
distributions of these states. A doorway
from the partonic to hadronic descriptions is the study of electromagnetic 
structure functions, and consequently
of PDFs, in the small-$x$ limit. Experiments at HERA \cite{hera}, and also at 
CERN \cite{cern} and the Tevatron \cite{fermilab}, 
have already begun to build a
bridge between hard and soft reactions, between the partonic and hadronic 
pictures of QCD. Much of the
intellectual excitement engendered by these developments comes 
from the realization that they afford a window
into a new regime of quantum field theory: high density, nonlinear 
interactions and yet weak coupling. And beyond
that, they link the short distance picture of QCD even further, to 
its phase structure; from the $Q^2-x$ plane of
momentum transfer and parton fraction \cite{levinth}, to the $T-B$ plane of 
temperature 
and baryon number density \cite{BT}. A few of the
important questions involved in this expansion of interest are given below.

\begin{itemize}

\item{} Saturation: Do parton densities at small $x$ access the 
high-density/weak-coupling
regime?

\item{} The QCD pomeron: What is the true high-energy behavior of 
hadron-hadron scattering?

\item{} Diffraction \& rapidity gaps: How can we understand the persistence 
of diffractive events in the presence of hard scattering? 

\item{} Colored glass condensate: Is it possible to compute bulk properties 
of 
the initial state in hadronic (nuclear) collisions?

\item{} Jet quenching: What is the behavior of high-energy partons traversing 
dense matter?

\item{} Quark-gluon plasma;
color superconductivity: What is the phase structure of QCD?

\end{itemize}

Even to a casual observer, these terms and questions relate to areas in which 
particle and
nuclear physics have influenced one another deeply over the past decade. 
The first three are perhaps more ``particle", the latter three, more 
``nuclear". 
Given the forum and constraints of space, we will concentrate on the former.
Nevertheless, an equally-important remaining question
is how to foster this interchange between the two fields \cite{capstick,bjorken}, 
given intellectual history 
and the realities of
funding.

\subsection{Small $x$ and Saturation}

In deep-inelastic scattering, $x=Q^2/2p\cdot q =Q^2/(s+Q^2)$,
with $p$ the
momentum of the target hadron, $q^2=-Q^2$ the momentum transfer through an 
electroweak current,
and $s\equiv (p+q)^2$.
The limit $x\rightarrow 0$ at fixed momentum transfer implies an
unlimited growth of center-of-mass energy in the target-current system
\cite{carli}. The limit $s\rightarrow \infty$
with momentum transfer fixed is the Regge limit, and the behavior of DIS 
structure functions as
$x\rightarrow 0$ is a species of total cross section, related by the optical 
theorem to the forward
scattering amplitude of an off-shell electroweak boson with the hadronic 
target.

In terms of the hadronic tensor
written as
a Fourier transform,
\ba
W^{\mu\nu}(x,Q^2)
={1\over 8\pi}\; \int d^4z\, {\rm e}^{-iq\cdot z}\; 
\langle p|{\cal J}^\mu(z)\, {\cal J}^\nu(0)|p\rangle\, ,
\ea
$x\ll 1$ translates to $z^2\sim 1/Q^2$, $p\cdot z \sim 1/x\gg 1$, and thus a 
correlation between
electroweak currents ${\cal J}$ separated by a distance of order 
$p\cdot z$ out along the light cone. As $x$ vanishes, this separation exceeds 
the size of the hadron,
and the measurements from which we determine PDFs take place not in the 
hadron proper, but in a region
where the hadron has dissolved into the surrounding vacuum. In this 
configuration, it may be better to
think of the PDF not so much as a measure of pre-existing partons within the 
hadron, as the probability
that a quark {\it pair}, a component of the nearby vacuum, polarized by the 
incident electroweak boson,
scatters from the target as a whole. In this picture, at large $s$, the 
electroweak boson
fluctuates into a virtual hadronic state that is typically off-shell by order 
$Q$, but whose (Lorentz
dilated) lifetime is long compared to its transit time across the target.

Rephrased in partonic language, at small $x$ the dynamics that determines 
the distribution of the quarks to which an electroweak boson couples is to be 
found, not inside the
target, but rather outside in the (target-independent) vacuum. At small 
impact parameter, the extended,
virtual hadronic states will interact with a nucleon with essentially unit 
probability, that is, with a
cross section determined by the geometric size of the target. 
This picture is often realized in terms of the perturbative coupling of a 
photon
to a dipole in the form of a quark pair, whose interaction with the target can
then be modeled \cite{dipole}.
When translated back to a PDF, this
implies saturation, a limit beyond which (in $x$) the distribution cannot 
grow. On the other hand, at
peripheral impact parameters, the scattering can be diffractive, 
transforming an incoming photon, for example, into recognizable hadronic 
states, such as vector
quarkonia. These can be thought of as messages from the QCD vacuum, carried 
by electroweak currents.

The qualitative pictures of partons within and without the hadron
have quantitative
analogs in complementary factorizations and evolutions, commonly 
referred to as DGLAP and BFKL. Both may be thought of as describing
the population of regions of phase space that open up in high energy 
scattering.

The classic factorization for DIS,
in terms of parton momentum fraction, leads, as we have described above, 
to an evolution in momentum transfer:
\ba
F(x,Q^2)=\int dz\, C(x/z,\mu)\, f(z,\mu) 
\Rightarrow \mu{\partial f(x,\mu)\over \partial \mu} = \int dz\, P(x/z)\, 
f(z,\mu)\, .
\ea
This is DGLAP evolution, and corrections to it are suppressed by powers of 
$\mu$
The evolution kernels, $P$,
describe the formation of virtual states through radiation processes that
decrease momentum fraction but increase transverse momentum. 
A similar relation may be written down for diffractive scattering at large 
momentum transfer, leading to the
concept of diffractive parton distributions
\cite{alve98}: the likelihood of striking a 
parton with momentum fraction $x$ at
a short-distance resolution $1/Q$ and, and still scattering the target 
elastically.

At small $x$, in contrast, the
relevant region of phase space is in rapidity ($y\sim\ln(1/x)$). The 
universality of the vacuum is
expressed as a factorization in rapidities, with a reference rapidity $y$, 
playing the role of the
factorization scale, $\mu$, above, and with a convolution in transverse 
momentum. Once again, the
factorization leads to evolution \cite{bfklevol}: 
\ba F(x,Q^2)=\int
d^2k_Td^2k'_T\, C(x,k_T-k_T',y)\, G(k_T',y) \Rightarrow {\partial 
G(k_T,y)\over \partial y} = \int
d^2k_T'\, K(k_T-k_T')\, G(k_T',y)\, . 
\ea 
This result, BFKL evolution, has, in general, corrections
that are power-suppressed in rapidity, rather than momentum transfer. Such 
corrections, for purely
practical reasons, tend to be rather more important, and harder to control. 
At the same
time, the two evolutions complement each other, and a great deal is to be 
learned by exploring what
each means for the other \cite{smallxevol}. The story of HERA small-$x$ data is in large part 
the recognition that
structure functions increase rapidly toward $x=0$, with a slope that 
increases with the momentum
transfer. The pendulum has swung from early enthusiasm for a BFKL 
explanation, to a realization that
the data even at moderate $Q^2$ are describable within the standard PDF-DGLAP 
formalism, back to a
realization that at $Q$'s below a few GeV, structure functions may be showing 
signs of power behavior,
and even saturation \cite{levinth}. 

\subsection{The Pomeron in 2001}

Through $x\rightarrow 0$, DIS, the quintessentially partonic process, 
leads back to the questions of hadronic scattering that were, 
some decades ago, displaced at the center of
strong interaction physics: how can we understand the behavior of the total 
cross section at high
energy? Is the Froissart bound saturated by the strong interactions, and if 
so, what is the dynamical
source of that saturation? What is the makeup of the total cross section in 
terms of diffractive and
highly inelastic scattering, and what does this tell us about hadronic 
structure? What is the
singularity structure of hadronic scattering amplitudes in the complex plane 
of angular momentum?

Beyond their undeniable historic interest, these questions highlight the
challenges in constructing a theory that makes the transition from
partonic to hadronic degrees of freedom. They are a set of questions 
that were, for a time, edged off center stage as we learned that rare, 
high-momentum events were a
window to QCD in terms of its most elemental degrees of freedom. The 
questions remained, however, and
we can recognize them now as a variant of one of the essential intellectual 
challenges: how complex
phenomena arise out of simple, underlying laws. 
The spirit of these questions may be summarized by the
technical query: what is the QCD pomeron \cite{landshoff}?

The pomeron is, to many of us, an
elusive concept, with origins in a Regge technology that is no longer widely 
accessible.  It can also be addressed in the very modern language of duality 
\cite{tan}.
In some sense, however, the concept is not a
difficult one.  It is an empirical
fact that most inelastic collisions result in energy loss through the 
production of particles with relatively low transverse momenta, and without 
strong correlations to the
initial-state hadrons. As initial-state energy increases, the phase space 
increases,
and the lack of correlations 
with the initial-state particles implies a universality for the bulk of the 
event. 
The optical theorem
-- that is, the conservation of probability -- identifies the total cross 
section
with the elastic scattering amplitude
at zero momentum transfer.
The pomeron can be thought of as the sum total of what is left behind when 
two hadrons collide, 
insofar as it is encoded into the elastic scattering amplitude. The 
universality
of central particle production
is essential here, and this universality ensures that the dominant behavior of
the DIS cross section is  related to the very same pomeron 
that
governs the total pp and $\rm p \bar p$
cross sections at high energy.
To the extent that elastic
scattering is thought of as resulting from the exchange of an object, that 
object
may be thought of as the pomeron. To identify it with a virtual particle, 
however, 
is certainly an oversimplification.  

The universality of the pomeron appears
not to be absolute, but to depend at least on the virtuality of the initial 
states 
to which it couples, or alternately whose
evolution into virtual states it describes. Thus, as $Q^2$ increases, the 
$x$, or
equivalently $s$, dependence of a PDF is expected to steepen, corresponding 
to 
the short-distance size of the initial state at the current end. Depending on 
the kinematics, the
actual increase may be determined by the DGLAP or BFKL equation, or either, 
or something else \cite{ccfm}. For a
pomeron ``pinned down" to short distances at both ends, it is possible that a 
really reliable
perturbative approach might be possible. An example will be afforded by the 
total cross section for
two virtual photons, with $S_{\gamma\gamma}\gg Q_1^2,Q_2^2\gg\Lambda_{\rm 
QCD}$. Such measurements,
already made at LEP, would be an exciting program for a future linear 
collider \cite{burr01,Abe:2001nq}. 

\subsection{Diffraction and Gaps}

One of the striking developments of Run I at the Tevatron was the 
observation of the diffractive production of heavy states, involving jets, 
electroweak bosons and
heavy quark pairs \cite{Goulianos}. 
At the percent level, a few high-$p_T$ processes stand like isolated 
islands above a
background of little or no radiation. At low overall momentum transfer, it 
appears natural to describe
this as the result of pomeron exchange between the hadrons. As in DIS, we may 
formulate, and to some
extent measure, diffractive parton distributions in hadron-hadron scattering, 
although they need not be,
and indeed are not, the same as in DIS. Other outstanding results from Run I 
involved dijet rapidity
gaps \cite{dijet,dijet630}, with little radiation between jet pairs 
resulting from large momentum transfers,
also seen at HERA \cite{ZEUS},
and the observation of intra- and interjet coherence effects in final states 
\cite{Fermicohere}.

The continued study of diffractive
and rapidity gap final states
at Run II of the Tevatron and at HERA is likely to be very fruitful, 
independently and in their comparison. Diffractive and gap events should be 
thought of as windows into
QCD evolution over long times \cite{diffrac}.
A gold mine of information on the transition of QCD dynamics from short
distances to long is encoded in correlations of radiation in diffractive 
final states, and their
analogs in minimum-bias \cite{field} and hard-scattering events.
The huge lever arm of the LHC, and of a
TeV-scale lepton beam, either in conjunction with a fixed hadronic target or 
a hadron beam,
could allow, if the concepts are in place, a precision study of QCD time 
evolution. This is an opportunity that will only be realized though a 
collaborative effort
of theory and experiment, beyond what we have accomplished in the recent 
past.

\subsection{Nuclei} 

High energy scattering on nuclei has led over the years to a 
series of surprises and challenges for QCD.   The Cronin effect
(enhancement of single-particle spectra at 
$p_T\sim$ few GeV), the EMC effect and shadowing at small $x$
(deviations from $f_{i/A}=Af_{i/N}$) were surprises at the time,
which led to deeper understanding of the meaning of parton distributions,
and of the role of multiple scattering at high energy.  
In factorized cross sections, nuclei differ from an
incoherent collection of nucleons in that their nonperturbative
distributions may -- indeed must -- differ, and also in that
power corrections in momentum transfer are enhanced at least
by a factor $A^{1/3}$, due to the longer path length in the nucleus.
This nuclear enhancement is characteristic of ``rare" hard scatterings,
in which the active projectile parton is unlikely to encounter more than a 
single
parton of the target.   In such cases, multiple scattering should be
thought of as a correction, of the generic magnitude $T_4 A^{1/3}/Q^2$,
where $T_4$ is a 
nuclear (higher twist) matrix element with mass dimension of momentum 
transfer squared \cite{luo92}.
In this case, nuclear effects give insight into long-range correlations
in the nucleus.

At small $x$, the difference between
nucleon and nuclear scattering can become much more pronounced.  
For example, as we
have observed above, the nucleus may interact not with 
a single parton, or electroweak boson, but with a long-lived
hadronic virtual state, often modelled by a QCD dipole.
For small impact parameter in this scattering, the interaction
can be geometrical, and grow as the surface area $\sim A^{2/3}$.
In this case, a nuclear target is expected to accentuate the transition from
evolution in transverse momentum to evolution in rapidity,
and consequent saturation, and the achievement of high-density
dynamics at weak coupling.

In the past decade, the discipline of nuclear physics has been transformed,
by the importation and further creation of ideas and methods in QCD.  The 
reason for these developments are complex, but there is no doubt that the
construction of the Relativistic Heavy Ion Accelerator at Brookhaven,
with the aim of observing a quark-gluon plasma, has had a tremendous 
impact.  As this project has matured, the shared interests of high
energy and nuclear physics have become more and more clear.
A keen appreciation of theoretical efforts in the
full range of QCD has characterized the RHIC project from the start,
and has stimulated ideas which are equally relevant to the high
energy program.  Prominent among these are the polarization studies,
which we will review below, but the full range of ``pA" and ``AB"
collisions planned at RHIC will pick up where
programs at the ISR and Tevatron left off.   One of the opportunities
and challenges
facing the high energy community is how to interact constructively
with this effort, how to benefit from and contribute to 
the study of QCD in a nuclear environment.   

\subsubsection{Nuclear Targets}

It is widely held that semihard processes play a increasingly 
important role in ultrarelativistic
heavy-ion collisions at $\sqrt{s}\geq 200$ GeV, in describing global
collision features, such as particle multiplicities and transverse energy
distributions.
These processes can be calculated in the framework of
perturbative QCD and they are crucial for determining the
initial condition for the possible formation of
a quark-gluon plasma.  To test the applicability of the
factorization theorems to these processes,
it is essential
to know the parton distributions in nuclei.  For example, to
calculate so-called ``minijet" production for $p_T\sim 2$ GeV,
 or charm
quark production
in the central rapidity region ($y=0$), one needs to know the nuclear
gluon distribution $xG_A(x,Q^2)$ at $Q\sim p_T$ or $m_c$ and
$x=x_{\rm Bj}=2p_T/\sqrt{s}$ or $2m_c/\sqrt{s}$. For $\sqrt{s}\geq 200$
GeV, the corresponding value of $x$ is $x\leq 10^{-2}$. Similarly, 
nuclear quark densities, $xf_{q/A}(x,Q^2)$, are needed for
the Drell-Yan lepton-pair production process.

The attenuation of the quark density in a nucleus has been firmly established
experimentally at CERN \cite{cern} and Fermilab \cite{fermilab}
in the region of small values of $x=Q^2/2m\nu$ in
DIS on nuclei, where
$Q^2$ is the four-momentum transfer squared, $\nu$ is the energy loss
of the lepton and $m$ is the nucleon mass.
 The data, taken over a
wide (but correlated) kinematic range, $10^{-5}<x<0.1$ and 0.05 ${\rm GeV}^2 <
Q^2< 100$ GeV$^2$, show a systematic reduction of the nuclear structure function
$F_2^A(x,Q^2)/A$ with respect to the free nucleon structure function
$F_2^N(x,Q^2)$. There are also some indications of nuclear gluon shadowing
from the analysis of J/$\Psi$ suppression in hadron-nucleus
experiments \cite{glushdw}. Unfortunately, the extraction of the nuclear
gluon density is not unambiguous, since it involves the evaluation of
initial parton energy loss
and final state interactions \cite{fs}.
These measurements present a tremendous challenge to the theoretical
models of nuclear shadowing (for recent reviews, see \cite{rev}).

The partonic description of nuclear shadowing has been
extensively investigated  in the context of 
recombination models \cite{glr,mq,eqw} as well as Glauber multiple scattering
theory \cite{glauber,frankfurt,lu,zakharov,mueller,agl}. 
These approaches predict
different scale ($Q^2$) dependence of the nuclear structure function
than DGLAP
evolution in transverse momentum space. Qualitatively, the scale evolution of 
a
nuclear structure function 
is slower than for a nucleon, leading to a perturbative mechanism
for the depletion of the parton density at large momentum
transfer.   The quantitative prediction of nuclear shadowing
at large $Q^2$ in general depends on the initial value of shadowing
at a small momentum transfer, where a perturbative calculation breaks
down and a non-perturbative model and/or experimental input is needed.
The so-called vector-meson-dominance (VMD) model \cite{vmd} has been 
successful in predicting the quark shadowing at small $Q^2$ where
the experimental data is rich \cite{cern,fermilab}.

\subsubsection{Small-x Gluon Distribution in a Nucleus}

Presently there is  rather little experimental
information on the initial gluon shadowing value at some small $Q_0^2$.
This, in principle, could make the prediction on gluon shadowing
very uncertain.  There exists, however, a typical scale
$Q_{\rm SH}^2\simeq 3$ GeV$^2$ in
the perturbative evolution of the nuclear
gluon density  beyond which the nuclear gluon shadowing can be
unambiguously predicted in the context of perturbative QCD \cite{hls}. 
This semihard scale $Q_{\rm SH}^2$ is determined by the
strength of the scaling violation of the nucleon gluon density in the
small-$x$ region, at which the anomalous dimension
$\gamma \equiv \partial \ln xg_N/\ln Q^2$ is of order unity. 
As $Q^2$ approaches $Q_{\rm SH}^2$,
the gluon shadowing ratio rapidly approaches a unique
perturbative value, determined  by Glauber multiple scattering theory.
At $Q^2>Q_{\rm SH}^2$,  the  nuclear
gluon density, in contrast to the quark case,
  is almost independent of the initial
distribution at $Q_0^2$, and its evolution is  mainly governed by 
DGLAP evolution.

\subsubsection{Nuclear Collisions}

The high parton density phase of QCD will play a central role in
high energy nucleus-nucleus collisions such those at the RHIC and LHC.
The primary purpose of these experiments is to create a
quark-gluon plasma, a stage through which our expanding universe must
have gone, and to investigate its properties. 
In this process, the state of high parton density is thought
of as an initial condition for the creation of a plasma,
resulting immediately after the collisions of the 
QCD fields of the initial-state nuclei.  Energy loss \cite{eloss}, 
resulting in so-called jet quenching, the softening
of high-$p_T$ spectra in nuclear collisions, may be a sensitive 
tool to study the evolving partonic state \cite{phenix}. 
Phenomena of this kind are typical of the current overlap of 
nuclear and high energy physics. 

To define the total number, $N$ of gluons in a hadron requires 
a resolution scale.  A natural intrinsic scale, which grows at small 
$x$, is 
\ba
\Lambda=\frac{1}{\pi R^2} \frac{dN}{dy}\, .
\ea
HERA data \cite{hera,carli} already suggests that $\Lambda>>\Lambda_{QCD}^2$, 
so that 
$\alpha_s(\Lambda)<<1$, {\it i.e.}, we are in the regime of weakly 
coupled QCD.   Furthermore, in case of a nucleus, the intrinsic scale, 
\ba
\frac{1}{\pi R^2} \frac{dN}{dy} \sim \frac{A^{1/3}}{x^\epsilon}\, , 
\ea 
and at small $x$ or large $A$, classical field methods may be applicable 
\cite{mv}.  
As $x$ decreases, $\Lambda$ increases,   
corresponding to larger $p_T$.  A typical parton size is of the 
order $1/p_T$.  Thus we increase the number of gluons by adding more gluons 
with smaller and smaller sizes, which cannot be seen by probes of 
size resolution $\geq 1/p_T$ (at fixed $Q^2$), and thus do 
not contribute to the cross section (and the cross section does not 
violate unitarity).  At the saturation momentum, $Q\sim Q_s$, 
the quantity $(1/\pi R^2Q^2)xG(x,Q)$ approaches unity, and
all powers of $Q$ (twists) become relevant.  The
saturation momentum depends on the 
hadron in question and on the longitudinal momentum fraction of the 
gluon.  At the RHIC and HERA, one encounters saturation scales of 
about 1 GeV, at the boundary of the perturbative region.  At the LHC, 
saturation scales in the 2-3 GeV range are anticipated.  

 One of the interesting theoretical developments in
connection with the RHIC project is 
an effective field theory approach for 
studying small $x$ physics in the regime of high density \cite{mmm}.  
The small 
$x$ effective action was obtained by successively integrating out 
the modes at larger values of $x$, which leaves the 
form of the action unchanged, while the weight satisfies a 
Wilsonian nonlinear renormalization group equation.  If the parton 
density is not too large, the RG equation can be linearized and it 
becomes BFKL equation.   Furthermore, in the double logarithmic 
approximation (small $x$, large $Q^2$), the renormalization
group equation can be expressed 
as a series in inverse powers of $Q^2$, where the leading term is 
the small $x$ DGLAP equation and the first subleading term corresponds to 
GLR equation.  
It is argued in Refs.\ \cite{mv,mmm} 
that the high parton densities in a
large and energetic nucleus at the RHIC 
and later the LHC, allow one to formulate the collision
classically, and to determine the initial conditions of such collisions
from QCD, and to predict
multi-particle and entropy production at
high energies from first principles.  The configuration of
gluon fields in such a picture has been picturesquely dubbed
the ``colored glass condensate" \cite{mclerran}.

A further proposal is to build an Electron-Ion Collider (EIC) at Brookhaven
National Laboratory. This would use the already existing heavy ion
beam from the RHIC, at $100$ GeV to collide with an electron beam of
$10$-$15$ GeV, which will achieve a center of mass energy of
$\sqrt{s} \sim 65$ GeV. This machine would expand the covered
$x$ and $Q^2$ kinematic region by an order of magnitude over
previous or current experiments such as NMC at CERN and E776 at Fermilab.
It would allow one to measure nuclear structure functions at very
small $x$ and moderate $Q^2$, the ideal region for investigating
high parton density effects using weak coupling methods.  It would
further shed light on the nature of nuclear shadowing and its $Q^2$
dependence at small $x$, which would be crucial for understanding
nuclear shadowing from QCD. For the first time, one would also be able
to measure the longitudinal structure function $F_L$ directly by tuning
the beam energy. $F_L$ is known to be a sensitive probe of high parton
density dynamics. The knowledge of nuclear structure functions
at small $x$ and moderate $Q^2$ would also be required in the upcoming
experiments at the LHC in its nuclear mode of operation, where parton
distribution functions at $x\sim 5 \times 10^{-4}$ at
$Q^2 \sim 1$ Gev$^2$ will be needed.

\section{Polarization}
\label{sec:polarization}

{\it This section was prepared by George Sterman and Werner Vogelsang.}

In polarization studies at high energy, we address 
the essence of what QCD has to offer: the experimental study of
strong coupling physics through degrees of
freedom that interact weakly.  More and more,
particle and nuclear physics facilities complement
each other in this area, as at the
HERA experiment, HERMES, and at the RHIC spin project at Brookhaven. 
The future development
of polarization physics in the high energy program will
depend in large part on a creative dialog between the
disciplines.  In the high energy community, however,
proton structure is  often thought of as relevant only
through its impact on
new physics searches at hadron colliders.  Perhaps it is worth noting
that the most ambitious programs for revealing large
extra dimensions, or black holes, involve the
same basic approach:  the study of strong coupling (string
theory) through the weakly interacting quanta of the Standard Model.
Should there come a time when arguments from duality
reach a sophistication that allows the construction of brane world
scenarios from M theory, the same methods will most likely be testable
by measuring the spin structure of the nucleon.

The past of polarization studies has been in deep-inelastic
scattering.  Its future will complement these studies with,
for the first time, hadronic collisions at RHIC~\cite{rhic}, starting
at the time of this writing.

\subsection{Deep-inelastic Scattering}

The analysis of polarized deep-inelastic scattering begins
with the hadronic tensor,
\begin{eqnarray}
{\cal W}^{\mu\nu} (P,q,S) &=&\frac{1}{4\pi} \int d^4 z \;
{\rm e}^{i q\cdot z}
\langle P,S|\left[ {\cal J}^{\mu}(z),{\cal J}^{\nu}(0) \right]
|P,S\rangle \nonumber \\[6mm]
&&\hspace*{-3cm}
=
{\cal W}^{\mu\nu}_{\rm unpol} (P,q,S)  +\; i
\,M\,\varepsilon^{\mu\nu\rho\sigma} q_{\rho} \Bigg[
\frac{{S_{\sigma}}}{P\cdot q}
{\; g_1 (x,Q^2)} + \frac{{S_{\sigma}} (P\cdot q)-
P_{\sigma} ({S}\cdot q)}{(P\cdot q)^2}
{\; g_2 (x,Q^2)} \Bigg] \; ,
\end{eqnarray}
where of the two new structure functions, $g_1$ is
the better-known, because it appears at leading
power in $Q$.  In addition, it has a convenient
parton-model interpretation, in terms of
polarized parton distributions,
${g_1 (x)} =
{\frac{1}{2} \sum_q e_q^2 \Big[ \Delta q(x) + \Delta
\bar{q}(x) \Big]}$ where
${\Delta f(x,\mu) \,\equiv\, f^{{+}}(x,\mu)-
f^{{-}}(x,\mu)}$ denotes the difference between densities of
partons with positive and negative helicity.
$g_1$ is accessible in experiments with longitudinally-polarized
targets, $g_2$ with transverse polarization.

The past decade has seen considerable progress in
the measurement~\cite{data} of $g_1$, so that roughly speaking this polarized
structure function is known about as well as $F_2$
was in the mid-eighties.  Evolution effects are well-observed,
and the difference between proton and neutron is well
enough known that the Bjorken sum rule~\cite{bj}, relating
  $\int dx(g_1^p(x)-g_1^n(x))$ to the axial charge $g_A$ that
governs the beta decay of the neutron,
is tested at around the ten percent level~\cite{data}.
As a result, there is now
an abundance of fits to polarized parton distributions~\cite{grsv},
which, however, are not always mutually consistent.
The range of these models underscores the importance
of a global approach to
polarized parton distributions.  The practical problem is
that the inclusive DIS
cross section only sees the gluon at next-to-leading
order.   This is its great strength
as a determination of quark distributions,
but leaves the gluon in the dark, as it were.

The measurement of $g_2(x)$, on the other hand, is
much more difficult than that of $g_1(x)$, because
$g_2$ only appears in terms that
are power-suppressed by factors of $m_p/Q$ in standard cross sections.
Nevertheless,
$g_2$ is a linear combination of leading-twist
polarized quark distributions, which depend on
matrix elements in the proton of the form $\langle p|\bar q q|p\rangle$, 
with $q$ a quark field, and 
higher-twist
quark-gluon correlations, $\langle p|\bar q F q|p\rangle$~\cite{vs},
with $F$ the gluon field strength.  
The recent precise measurements~\cite{e155} thus shed light
on the importance of these elusive windows into
nucleon structure.

\subsection{Hadronic Beams and $\Delta G$}

The limitations of DIS for determining the
polarized gluon distribution have been
a large part of the impetus
for a new generation of polarized beams of hadrons.
The spin program at  RHIC~\cite{rhic} will provide, for the first
time, colliding beams of both longitudinal (and
%(eventually) 
transverse) polarization, which will
bring into reach processes such as high-$p_T$
direct photon production, for which polarized
gluon distributions appear at leading order.
If the momentum transfer is large, these cross sections
have the standard factorized form,
\begin{eqnarray}
\hspace*{0cm}d {\Delta} \sigma_{AB}^{{\gamma}}
&=& \sum_{a,b} \int dx_a\, dx_b \,\,
{\Delta}f_a (x_a,\mu) \, {\Delta}f_b (x_b,\mu)\;
d{\Delta}\hat{\sigma}_{ab}^{{\gamma}} (x_a\, P_A\,,x_b
\,P_B\,,p_T^{\gamma},\eta^{\gamma},\mu)\, ,
\label{hhpol}
\end{eqnarray}
with $\Delta\sigma$
the helicity-dependence of the short-distance
cross section,
$d{\Delta} \hat{\sigma}_{ab}^{\gamma}\;=\;
(1/2)\,[ d\hat{\sigma}_{a(+)b(+)}^{\gamma}\;-\;
d\hat{\sigma}_{a(+)b(-)}^{\gamma}]$.
As in the case of unpolarized distributions, the eventual
determination of polarized PDFs will require more than
a single input, and hadron + hadron $\rightarrow \gamma+X$
will be complemented by other channels in proton-proton 
scattering, such as jet or heavy flavor production.
A promising example
is also charm photoproduction~\cite{gr,compass,e161}, for which the gluon 
fusion 
processes,$\gamma g\rightarrow c\bar c$, is relevant at leading order
in $\alpha_s$. In measurements
of {\em semi-inclusive} DIS, such as those already performed by
SMC~\cite{smc} and HERMES~\cite{hermes},
one looks for specific hadronic final states and can tag
individual polarized quark flavor distributions.
Similarly, the  RHIC experiments can
achieve this 
through W boson production~\cite{rhic}.  In the not-too distant
future, we will have available the information necessary
for a truly global fit for parton helicity distributions~\cite{sv},
hopefully with quantifiable uncertainties.

\subsection{Beyond Helicity}

The experiments discussed so far are all aimed
at the helicity distributions $\Delta f$ of the nucleon,
the difference between parton densities of flavor $f$ with
helicities along and against the helicity of the parent hadron.
There is much to learn, however, from transverse
polarization as well.
Historically, fixed target {\em single}-spin
experiments studied transverse
polarization in the initial state~\cite{single}
and hyperon spins in the final state~\cite{lambda}.
These experiments often showed striking spin
dependence, which
may be interpreted in partonic language
in terms of higher-twist quark-gluon correlations~\cite{qs},
of roughly the same form as encountered in $g_2$,
$\langle p|\bar q F q |p\rangle$.  Nevertheless,
the limited kinematic range available to these
older experiments make this interpretation far from conclusive.
The capability of the Tevatron in mapping these higher-twist
effects has, unfortunately, not been exploited.

Even at leading twist,
however, there is another matrix element accessible
in polarized hadronic scattering, the {\em transversity}~\cite{ralst}.
Transversity measures the interference between
states in which a parton 
carries opposite helicity, which can be nonzero
because of the spontaneous breaking of
chiral symmetry in QCD, and to a lesser extent
because the quarks are massive.
Defining projections
for  helicity and transverse  spin of
quarks by
${{\cal P}^{\pm}\;\equiv\;
({1\!\!\!1\pm
\gamma_5})/{2}}$ and
${{\cal P}^{\uparrow\downarrow}\;\equiv\;
({1\!\!\!1\pm
\gamma_{\perp}\gamma_5})/{2}}$, respectively,
the helicity and transversity distributions
can be put into similar forms,
\begin{eqnarray}
&&{\Delta q,\delta q} \left(x\right)\;\sim\;\frac{1}{2} \sum_X\;
\delta (P_X^+ - (1-x) P^+)\;\left[
\left| \langle X| \,{{\cal P}^{+,\uparrow}}q(0) \,|
P;\lambda,S_{\perp} \rangle \right|^2
{\mbox{{\bf --}}}\;
\left| \langle X| \,{{\cal P}^{-,\downarrow}}q(0) \,|
P;\lambda,S_{\perp} \rangle \right|^2 \right]\, ,
\end{eqnarray}
where the helicity
$\lambda$ and the transverse spin $s_\perp$
are 1/2 in either matrix element.
In inclusive DIS,
the interference terms decouple from
the hard scattering, except for the
small helicity-mixing effects of
quark masses.   In principle, the measurement of transversity
is a straightforward application of Eq.\ (\ref{hhpol})
for the hard-scattering
of two transversely-polarized colliding beams \cite{rhic}.
In practice, this is also
difficult, primarily because the hard-scattering
functions $\Delta \sigma$ that are
sensitive to transversity turn out to be
disappointingly small~\cite{js}.

The most promising window into transversity appears
to be by matching its helicity mixing with
corresponding effects in fragmentation.  Here,
the most popular observable is the ``Collins effect"~\cite{coll},
which depends on the observation that the
nonperturbative joint
distribution $D^h (z,\vec{k}_{\perp})$, with $k_\perp$ the hadronic
transverse momentum  relative to the axis of a jet, factorizes from the 
hard scattering
in the same fashion as do distributions in momentum fraction $z$ alone.
It is then possible that we may observe
nonzero values for such correlations as
\begin{equation}
\tilde{D}^h(z) = \int d^2 k_{\perp}\; D^h (z,\vec{k}_{\perp})\cos(\phi)\, ,
\end{equation}
where 
$\cos(\phi) ={\vec{k}_{\perp}\cdot \vec{w}}/{|\vec{k}_{\perp}|
|\vec{w}|}$, with ${\vec{w}\equiv \vec{S}_T \times \vec{k'}}$
in terms of the momentum $\vec k'$ of the outgoing electron.
Within the past year, hints of this effect have been 
reported~\cite{hermes1}, although the energies and momentum transfers
involved leave its interpretation in partonic
language still problematic.  A dedicated transverse-spin program at the
RHIC can, in principle, provide the lever arm in energy
to explore these, and related~\cite{others}, signs of helicity
mixing in hadronic structure.

Much of the current interest in polarized scattering resulted
from the the so-called  spin crisis, 
which boils down to the observation
that the total helicity of the proton appears
not to be carried primarily by quarks.  While this came
as something of a surprise, given the general success
of partonic language in the description of DIS, the
identification of nucleon with parton helicity is
in no way a prediction of QCD, perturbative or otherwise.
Nevertheless, if we must look elsewhere for the proton's
spin, orbital effects are the natural choice.  Formally,
one has in mind operator expectations
of the angular momentum tensor~\cite{orb,orb1},
\begin{equation}
1/2 \;=\; \langle P,1/2\,|\,J_3\,|\,P,
1/2\rangle
\;=\;\langle P,1/2\,|\,\int d^3 x\,M_{012}(x)\,|
\,P,1/2\rangle\, .
\end{equation}
The analysis of such matrix elements has led to
the introduction of a wider class of distribution,
the so-called off-diagonal or skewed parton distributions~\cite{orb1,orb2},
which take the general form
$\langle p+\Delta| \bar q q|p\rangle
$,
in terms of some momentum transfer $\Delta$.  Matrix
elements of this form interpolate between
DIS structure functions and elastic form factors,
as well as diffractive production amplitudes,
through the unitarity properties of the theory.
They are measurable in principle through ``deeply
virtual Compton scattering"~\cite{orb1}: $\gamma^* p \rightarrow \gamma p$,
which has been observed both at HERA and Jlab
quite recently~\cite{dvcs}.  We are still far from the quantitative
experimental surveys of DVCS and related processes
that would allow us to work backwards to
new insights into off-diagonal matrix elements
and angular momentum.  Nevertheless, a
direction has been set.

\subsection{At the Boundary of Nuclear and Particle Physics}

The proton is the simplest nucleus, and nucleon substructure
has become more and more a subject of investigation
for the nuclear community.  When it
comes to polarization studies in particular,
it is not possible to make a clean distinction between
the viewpoints of the two disciplines.
This should be thought of as a positive development,
and indeed a reprise of the heritage of
high energy physics.  In some
areas, particularly the study of off-diagonal
matrix elements, the initiative is more on the
side of nuclear physics.  Others, such as the
study of DIS distributions, seem more like particle
physics.  
Machines now under considerations
again know no boundary between 
specialties, including
an electron-ion collider (EIC \cite{eic}) with, like the RHIC, a
polarization capability, and a polarized fixed-target
option for TESLA (TESLA-N \cite{teslan} ) or another TeV-scale
linear collider.
In our view, this overlap of
specializations within shared interests constitutes an opportunity
for future collaboration.

\section{Programs for the Coming Decades}
\label{sec:programs}

Whither QCD?   The rules of quantum mechanics ensure that
quantum chromodynamics is a nearly ubiquitous 
feature in high energy physics, a key player in
nearly every experimental program, present
and projected.  

Among current colliders, the importance of
determining parton distributions, and quantifying
their uncertainties has been described above.  
We also emphasize the desirability of archiving
data in a form that is amenable to further analysis,
if possible allowing for the application of the
novel ideas and new methods that are sure to come.
Lattice QCD shares many methods and problems with 
perturbation theory, and we have seen
how it may interact with it, as with experiment, in the coming years.
Initiatives such as the SciDAC program will be needed
to ensure that improved computing capabilities will be
available to exploit theoretical advances in this
direction.

Although neutrinos do not themselves interact strongly,
their inelastic and elastic scattering on hadrons,
which dominate their tiny cross sections,
are both a source of information about hadronic structure,
and a source of uncertainty in the interpretations
of oscillation and related experiments.  
QCD will therefore play a crucial role in
the analysis of both cosmic and accelerator-based
neutrino experiments \cite{E1wg}, whose full potential will
be limited in some cases by strong interaction theory.
The atmospheric, water and ice cross sections
of ultrahigh energy neutrinos \cite{gandhi00} are, in principle, 
sensitive to hitherto unexplored ranges in parton
momentum fraction.  

The continued planning for,
and data analysis at, the Large Hadron Collider
will depend on the very parton distributions being
hammered out now.  The LHC itself will provide a
new frontier for QCD in a variety of ways,
by probing nucleons at smaller scales \cite{atlas01}, offering
new tests of quark compositeness and of the theory
of hadronization, extended ranges for diffractive 
and rapidity gap physics, windows into the
evolution of color degrees of freedom from
partonic to hadronic quantum numbers.

The Large Hadron Collider will also realize
in a single facility the shared goals of particle
and nuclear physics \cite{alice}, on the investigation of new
phases of QCD, and on their connection to the
realm of low parton-$x$, whose study
began in earnest at HERA.  

Looking forward to a new linear collider, a vast
field of QCD investigations opens up \cite{Abe:2001nq}.  Top pair production
allows studies of QCD at the electroweak symmetry breaking
scale, and a realization of the exquisite physics
of production thresholds.   
Photon-photon cross sections at a linear collider
will offer the first entry into forward scattering
in a perturbative region, and a direct test of
the celebrated BFKL program.  
Precision top quark physics will become a reality, with
implications for both Standard Model and post-Standard Model
physics.
A linear collider
will offer jet fragmentation over unprecedented
ranges, and a lever arm for event shape and energy
flow analysis that will realize the promise of studies at
LEPII.  With higher-loop calculations in hand,
and an improved theory of power corrections, it 
should be possible to measure the strong coupling
to the one percent level, which will offer a 
searchlight toward the physics of unification.
Studies of QCD at a linear collider will
both benefit from and strengthen QCD, and hence
searches for new physics, at hadronic colliders,
both the LHC and a VLHC or other successor.

Finally, should supersymmetry be discovered, from
Fermilab on up, a new world of strongly interacting
particles will emerge.  New problems, and new
rewards will flow from studies of QCD in this even
larger context.  

While it is notoriously difficult to predict
new directions in theory, it seems safe to predict
substantial progress in the following directions,
driven by the capabilities of experiment:

\begin{itemize}

\item{} Next-to-next-to-leading order
phenomenology of jet cross sections, with the aim of percent-level
precision.

\item{} Parton distribution uncertainties.

\item{} Next-to-leading order event generators.

\item{} Unquenched lattice calculations at the percent level.

\item{} General analysis of power corrections in infrared-safe 
cross sections, heavy quark decays and hard-scattering functions.

\item{} Theories of QCD energy flow in jet events and diffraction in hard scattering.

\item{} Behavior of the total cross section at high energy.

\item{} Elucidation of the polarization structure in the nucleon.

\item{} High density and high temperature QCD in nuclear collisions.

\end{itemize}

Progress on some of these projects, particularly the first two,
is well underway in the high energy community.  
In some others, progress will come increasingly from
the interplay of high energy and nuclear experiment and theory.
In yet others, 
only halting first steps have been taken.   A theory of
hadronization, and eventually of confinement is sure to come some day,
perhaps, as has often been suggested, out of string theory.  In any event,
there will certainly be other important steps forward.  The most
important of these will help to  bridge further the gaps between the
languages of quantum chromodynamics.

\section{Concluding Sentiments}
\label{sec:sent}

Quantum chromodynamics is an essential ingredient
in the future of particle physics on the basis of its intrinsic interest.
As a theory with weak and strong coupling phenomena, it is an
inexhaustible testing-ground for advances in quantum field,
and string theory, and as a component of the Standard Model
its exploration is basic science at its best.  

If there is a concern for the future of QCD in high energy physics, it is
that it is sometimes taken for granted.  It is common enough
to hear, in particular, that perturbative QCD is well-understood,
and that nonperturbative is synonymous with progress, 
which leads to the idea that the remaining frontiers of QCD are
irrelevant to high energy physics, and vice-versa.  
As we have stressed above, every experiment that involves
hadrons (even when only virtual, {\it viz.}\ muon $g-2$
\cite{lightbl}) is senstive to QCD dynamics at all
length scales.  The beauty of QCD is to be found in
its wealth of observables, its ``infinite variety",
that can in principle test every
concept of quantum field theory.  We would like to 
suggest that many of these concepts may not yet be
developed, but will take form as we ask new questions
of this inexhaustible theory.  
We suggest to both theorists and experimentalists
to take QCD seriously as an open
field of inquiry in its own right, to return to its basics,
to realize that much theoretical formalism was built up in
a different environment, before the data revolution of 
the nineties.  
Old ideas should be reevaluated and reworked 
when necessary.  Some aspects of theory must be
reinvented for new experimental realities, as 
can be seen now at the B factories,
or in polarization programs.  

The boundary between QCD for its own sake, and QCD as a servant
of new physics is an open one.  Studies in QCD and searches for
new physics will go hand-in-hand in the coming decade.  Indeed,
the complexity of new physics signals, documented again and again
at this conference, challenges us to develop our control over
QCD evolution at a level beyond what we have today.  
In particular, a 
TeV-scale linear collider program concurrent with the Large Hadron
Collider is a natural match for QCD studies, and for the
capability of QCD analysis to contribute in the search for
new physics.  To accomplish this development requires a broad, vigorous 
program
in high energy physics in the intervening time, with
dynamic interplay between theory and experiment
at the Tevatron, HERA, the LHC along with the
RHIC at Brookhaven and the
electron accelerator at Jlab.     It will
need coherent long-range planning that incorporates an appreciation
of current projects and facilities, as well as their support
in theory.  

\subsection*{Acknowledgements}

The conveners and authors of the working group
on QCD and Strong Interactions thank the Organizing Committees of
Snowmass 2001 for their tireless work, and for the many invitations,
reminders and encouragements that made the meeting and this
report possible.  We thank as well the conveners of many other
working groups for their collegial participation in many of
our meetings.
This work was supported in part by the National
Science Foundation and the Department of Energy.

%%%%%%%%%%%%%%%%%%%%%%%%%%%%%%%%%%%%%%%%%%%%%%%%%%%%%%%%%%%%%%%%%%%%%%%%%%

\end{document}